# A multiferroic on the brink: uncovering the nuances of strain-induced transitions in $BiFeO_3$

D. Sando[1*], Bin Xu[2], L. Bellaiche[2], and V. Nagarajan[1**]

[1]School of Materials Science and Engineering, University of New South Wales, Sydney 2052, Australia

[2]Physics Department and Institute for Nanoscience and Engineering, University of Arkansas, Fayetteville, AK 72701, USA

***daniel.sando@unsw.edu.au, **nagarajan@unsw.edu.au**

## Abstract

Bismuth ferrite ($BiFeO_3$) is one of the very few known single-phase multiferroic materials. While the bulk compound is rhombohedral (R), the discovery of an epitaxial strain-induced structural transition into a so-called 'super tetragonal-phase' (T-phase) in this material incited a flurry of research activity focused on gaining an understanding of this phase transition and its possible functionalities. This metastable phase of $BiFeO_3$ is also multiferroic, with giant ferroelectric polarization and coexisting antiferromagnetic order, but above all it is the strain relaxation-induced phase mixtures and their outstanding piezoelectric and magnetoelectric responses which continue to intrigue and motivate the physicist and materials scientist communities. Here, we review the research into the T-phase and mixed-phase $BiFeO_3$ system. We begin with a brief summary of the history of the T-phase and an analysis of the structure of the various phases reported in the literature. We then address important questions regarding the symmetry and octahedral rotation patterns and the (as yet underexplored) important role of chemistry in the formation of the metastable T-phase. We follow by describing the phase transitions in this material, and how these may hold promise for large magnetoelectric responses. Finally we point out some experimental challenges inherent to the study of such a system, and potential pathways for how they may be overcome. It is our intention with this work to highlight important issues that, in our opinion, should be carefully considered by the community in order to use this fascinating materials system for a new paradigm of functionality.

## 1. Introduction

The enhanced electromechanical response observed in the vicinity of a morphotropic phase boundary (MPB) in ferroelectrics is a critical phenomenon used extensively in piezoelectric applications.[1–3] At an MPB, crystallographically-different phases separated by low energy barriers coexist, giving rise to large physical responses (e.g., strain $\varepsilon$ and polarization $P$) to weak external stimuli (e.g., stress $\sigma$ and electric field $E$) (Refs. [4,5]). One can refer to materials tuned to be at such boundaries '*on the brink*', since these compositions can be considered as teetering between two strongly different states. The property enhancement arises from the flattening of the free energy profile across this boundary, meaning that even a slight external perturbation is sufficient to bring about a phase transition. The dissimilar physical properties of the two states results in giant responses when this phase transformation is induced.

The most common method for realising MPBs is through solid-state chemistry routes, via solid solution mixtures, of which the classic examples are $PbZrO_3$-$PbTiO_3$ (PZT) and $PbMg_{1/3}Nb_{2/3}O_3$-$PbTiO_3$ (PMN-PT), with the more recent (1-$x$)$BiTi_{(1-y)/2}Fe_yMg_{(1-y)/2}O_3$–($x$)-$CaTiO_3$ (BMTF-CTO) (Ref. [6]). These are therefore composition-driven and result from polarization-lattice interactions (predominantly through the oxygen sublattice) between competing phases.

On the other hand, physical handles can also be used to induce such 'on the brink' phases. This is achieved generally through external stress, although straddling an MPB in temperature space is also possible. In bulk ceramics this stress is typically applied via hydrostatic pressure, with some early experiments carried out as far back as the 1950s (Ref. [7]). More recently, Ahart *et al*. used hydrostatic pressure to induce, and gain insight into the origin of, an MPB in ferroelectric lead titanate ($PbTiO_3$ – PTO) (Ref. [5]).

In thin films, one can realise extremely large magnitudes of applied stress (on the order of several GPa) via epitaxial misfit strain; this is the so-called "strain-driven" approach. Strain-driven phase transitions in ferroelectric thin films have been reported in, for instance, various titanates: $BaTiO_3$ (BTO) (Ref. [8]), $SrTiO_3$ (STO) (Ref. [9]), and $PbTiO_3$ (PTO) (Ref. [10]). In this review, we focus on the epitaxial strain-driven phase transitions of the popular multiferroic compound bismuth ferrite ($BiFeO_3$ – BFO) (Refs. [11,12]). As observed for simple $ABO_3$-type ferroelectric thin films such as BTO (Ref. [8]) and PZT (Ref. [13]), BFO exhibits remarkable strain-induced modifications of its physical characteristics[14]. The multifunctional nature of BFO entails that the strain-driven transition offers a very unique opportunity in which not only coupling between the polarization and the lattice is modified (as in typical ferroelectrics such as PTO), but additionally, through the magnetic character of BFO, *spin-lattice* coupling can be explored through these strain engineering techniques. In addition, the intimate link between ferroelectric (FE) and the antiferrodistortive (AFD) degrees of freedom in BFO sets it apart from other typical perovskite materials: in most other $ABO_3$ compounds the FE and AFD degrees of freedom are typically mutually exclusive (*e.g.* in BTO there is FE but no AFD; while $CaTiO_3$ has AFD but no FE distortion), while in BFO both distortions coexist, and their interplay gives rise to remarkable phase diagrams[15] and physical properties[16,17].

### 1.1 History of the strain-induced phase transformation in $BiFeO_3$

The simple perovskite BFO crystallises in the rhombohedral $R3c$ space group, a structure which allows a polar distortion along the [111] pseudocubic direction – which appears at temperatures below 1100 K – breaking symmetry and giving rise to ferroelectricity[18]. This space group also allows antiphase octahedral rotations about the [111] direction, denoted $a^-a^-a^-$ in Glazer notation[19]. Furthermore, this material possesses long-range magnetic order with G-type antiferromagnetism below the Néel temperature of 640 K (Ref. [20]), with an interesting long-period spin cycloid[21]. Importantly, the bulk pseudocubic lattice constant is 3.97 Å, placing it in prime location for the multitude of substrates currently available for oxide thin film growth. Since BFO is also innately lead-free, it has been investigated as a prime candidate for environmentally friendly piezo-ceramics. While the bulk material had attracted some interest more than four decades ago, it was only after the seminal demonstration of exceptional physical properties in BFO epitaxial thin films[22] that this material initiated an immense surge of research interest into multiferroic materials[23,24], particularly in thin films.

When epitaxial BFO thin films are stabilised on substrates that impose a strong compressive strain[a], a striking example of the richness of strain-induced phase transitions can occur. Under such conditions, the material can crystallise as a metastable polymorph with a giant axial ratio ($c/a$ ~ 1.23), a phase which has come to be known as the *T phase* or T-like phase (T' phase) of BFO (Fig. 1c), to distinguish it from the rhombohedral (R) bulk parent compound (Fig. 1a). The notion that BFO could form in a "super tetragonal" phase possessing giant polarization was first explored theoretically in the mid 2000's by various groups[25–27]. Indeed, preliminary ferroelectric characterisation of intriguing 'mixed phase' films by Yun *et al.* in 2004 alluded to a very large polarization value of 150 μC/cm$^2$ (Ref. [28]). While an adequate explanation for these results was lacking in the initial report, subsequent experimental characterisation[29] appeared to corroborate the possibility of the T' phase fraction in the sample being the origin of the large polarization. The first report of the epitaxial stabilisation of this giant-axial ratio phase in its *pure* form by Béa *et al.* showed unequivocally that the phase was multiferroic[30], but somewhat surprisingly, the measured polarisation was not strongly enhanced relative to the bulk-derived R-like (R') phase. To this day, no group has reported a traditional ferroelectric hysteresis loop showing the predicted[25] 150 μC/cm$^2$ polarization for a *pure* T' phase BFO sample[b].

Several months after the initial experimental stabilisation of T' BFO on LAO substrates, the group of Ramesh (Berkeley) reported that T' films grown beyond the critical thickness for strain relaxation exhibited intriguing piezoelectric characteristics, and the breakthrough report of Zeches *et al.*[31] discussed in this context a 'morphotropic phase boundary' induced by strain alone[c]. It was shown in this work that an electric field, applied through a scanning probe microscope tip, could be used to convert regions of mixed R' and T' phases to pure T' phase (and back), giving rise to a very large strain, manifest as physical height changes in the sample. There was immediate excitement from the possibility of generating very large piezoresponses from this mixed-phase system.

The work of Zeches *et al.* opened up the exciting possibility exploiting electric-field driven (*E*-driven) phase transitions to create functional devices, and this prospect continues to motivate strong interest in this system. One such idea is the electric control of magnetoelectric coupling and/or magnetism, particularly at the phase boundaries. The mixed-phase of BFO exhibits intriguing magnetic properties: while the 'bulk' of T' and R' BFO are antiferromagnets[20,30] (along with, in the R' phase, a small moment due to spin canting[32–34]), photoemission electron microscopy (PEEM) measurements on mixed phase films indicate that locally a magnetic moment is present in the highly-distorted R' phase regions[35]. If one could use an electric field to manipulate these R' regions, then the associated magnetic moments could also be modified, bring *E*-driven magnetism one step closer. Another exciting possibility is to use the *E*-driven transition to induce large changes in optical absorption, exploiting the strong variation in the electronic structure of the different phases[36]. This property could enable new thin-film devices with electrochromic functionality.

---

[a] Strictly speaking, strong compressive strain is not a necessary condition for stabilising this phase. As highlighted later, other methods, such as changing growth conditions, can be employed.

[b] Substitution with 40% Ga yielded a giant axial ratio phase, and hysteresis loops with 150 μC/cm$^2$ polarization were measured on these films at room temperature recently[132].

[c] What defines an MPB, in a canonical sense, is still debated. Of course, MPBs are by definition formed when solid solutions create intricate phase mixtures, and are defined by a vertical phase boundary, so the labelling of this strain-driven transformation as an MPB is somewhat a misnomer. This is particularly evident upon inspection of the phase diagram in the supplementary material of Ref. [31]. However, the similarities between the strain-driven BFO and the solid solution systems probably justify this classification.

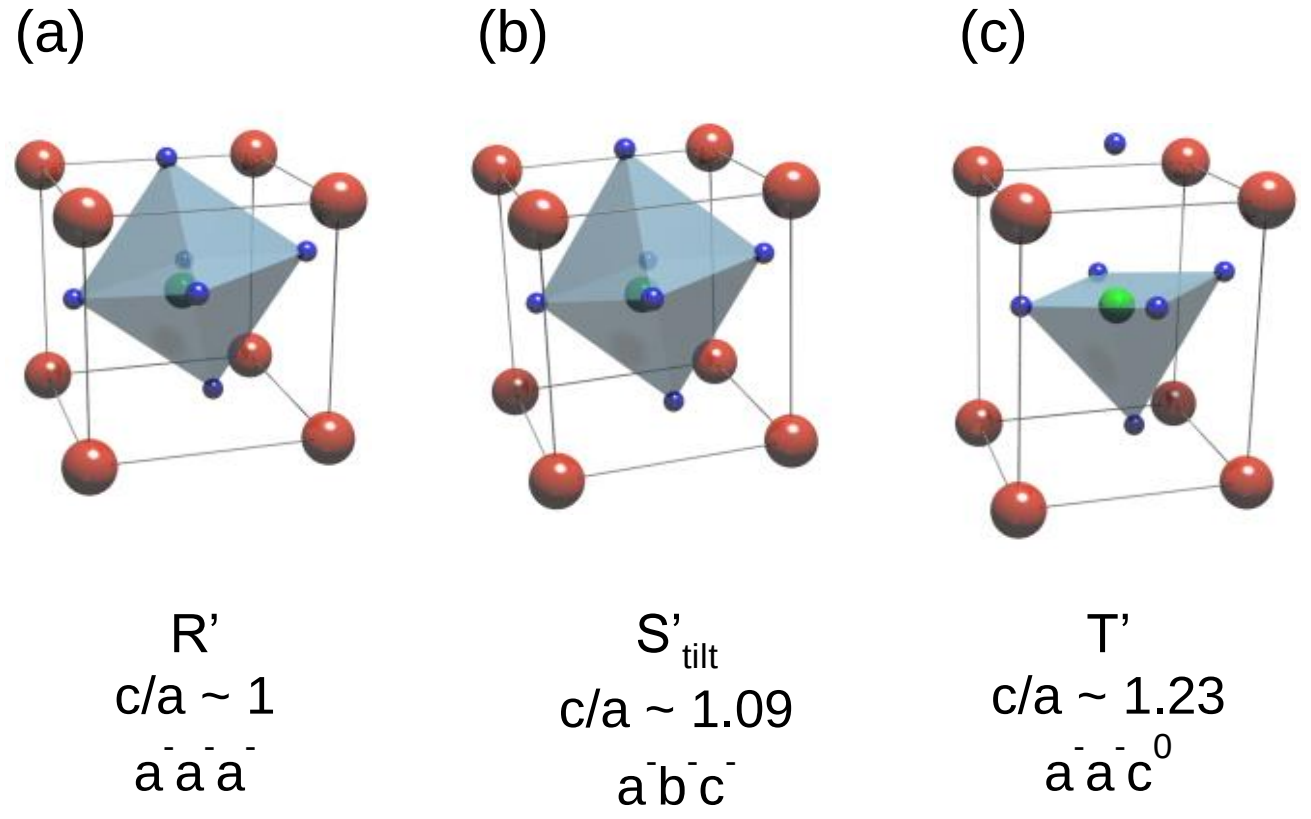

Figure 1. Structural variants in $BiFeO_3$ films, represented as pseudocubic unit cells. (a) The rhombohedral-like R' phase with octahedral coordination of the oxygen ions and equal antiphase oxygen rotations around all three axes. (b) The intermediate tilted S' phase with octahedral coordination and unequal antiphase oxygen rotations around all three axes. (c). The tetragonal-like T' phase with square-pyramidal coordination and antiphase oxygen rotations around the *a* and *b* axes only.

## 1.2 Motivation for the present review

We have taken the opportunity to write this review and consider it a timely contribution, because in the ~6 years since the experimental discovery of the T' phase, the literature is still compact enough (~120 publications) to treat fairly comprehensively. The past six years have seen intense research activity (*cf*. publication rates presented in Fig. 2a) which has delved into the intricacies of this rather complex system. A number of pertinent points may be emphasised from Fig. 2. First, as can be expected, the initial work (2005-2009) was for the most part theoretical and predictive in nature. The bulk of the publications in the 1-2 years after experimental discovery was focused on characterisation of the physical properties (structure, optical response, switching mechanisms) of the T' and mixed phases. A number of temperature-driven (*T*-driven) structural phase transitions were reported in 2010 and 2011. These observations in particular further piqued the interest of the ferroelectrics community, since the *T*-driven symmetry change ($M_C \rightarrow M_A$ upon increasing temperature), and associated polarization rotation, once again shows strong parallels with the MPB compositions in lead-based compounds. Toward the end of 2011–2012, the *T*-driven magnetic phase transitions received considerable attention[37–39], with two groups suggesting a coupling of magnetic and ferroelectric transitions[38,39]. This concept in particular is distinct to the strain-induced MPB of BFO: unlike typical ferroelectric oxides, this system holds the potential for *multiferroic* phase transitions. A final pertinent point from Fig. 2b is that publications dealing with the structural characterisation of the BFO T' or mixed-phase system have not waned over the years, despite the high intensity of research. This implies that the system is crystallographically-speaking remarkably complex and that open questions remain.

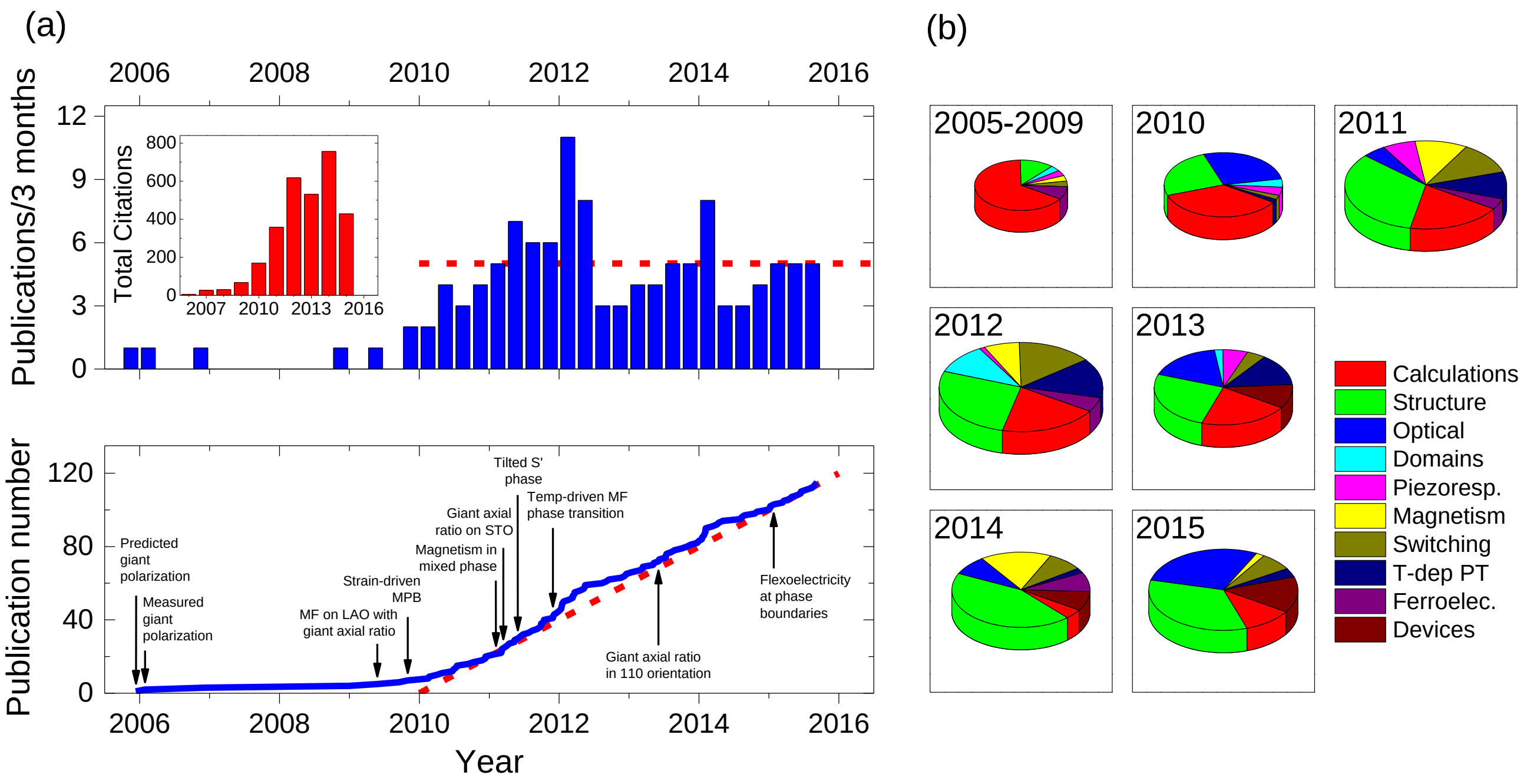

Figure 2. Chronology of tetragonal-like $BiFeO_3$ research. (a) Number of publications and total citations per three month period (top and inset respectively), and timeline showing publication progress and notable advances (bottom). The red dashed lines denote a publication rate of 20 papers/year. MF = multiferroic; LAO = $LaAlO_3$; MPB = morphotropic phase boundary; STO = $SrTiO_3$. (b) Breakdown of publications according to topic. The area of each pie is proportional to the number of publications in the corresponding period. Piezoresp. = piezoresponse, T-dep PT = temperature-dependent phase transitions; Ferroelec. = ferroelectricity.

---

We begin our review by presenting a complete summary of the reported structural characteristics of the various phases, having made the effort to carefully scan the literature and collate almost all available published structural data (Table I). Whilst the information contained in this table in itself is extremely helpful for anyone either beginning in this field or wishing to quickly compare results against published literature, our analysis has allowed us to devise the first ever *strain – growth rate*/*temperature* phase diagram for BFO thin films. This will be particularly useful for thin films growers interested in this system. Second, through this comprehensive data analysis we calculate the Poisson ratio of R' phase BFO, a useful physical characteristic for the study of thin films, using the largest dataset considered to date for this purpose. The Poisson ratio obtained here can be used in numerous thin-film based problems, ranging from the calculation of electro-mechanical coefficients to the single ion anisotropy. Importantly, during our analysis, we noticed that despite the intense experimental and theoretical exploration of this system, a number of vital – and somewhat obvious – questions remain and, more importantly, there are contradictory findings. Our aim here is to address these issues and, through a balanced survey of the literature, to raise key questions and provide perspectives on shortcomings where attention needs to be focused.

While it would be impossible to consider every characteristic of this system, we have identified four key issues which we believe merit further investigation: a) the **structure** of the various phases, in particular the **strongly-strained** **rhombohedral-like phase**, and its connection with diffuse peaks in XRD which thus far have received no attention; b) the **symmetry and octahedral rotation patterns**; c) the **role of chemistry**; and d) **phase transitions**. We conclude the review with a section dealing with some of the **experimental challenges** inherent to this system. Our intention is to stimulate and challenge the community, by bringing these issues to the fore, and suggesting potential pathways for the next iteration of investigations into this intriguing, multifaceted, and rather amazing materials system.

**Table I**. Reported structural parameters for T' and mixed-phase $BiFO_3$ films, from experiments and first-principles calculations. Parameters for the secondary phase(s) (if present) appear in the subsequent row(s) of a particular entry. Substrates: STO - $SrTiO_3$; LAO - $LaAlO_3$; YAO - $YAlO_3$; LSAO – $LaSrAlO_4$; NGO - $NdGaO_3$; LSAT – $(La,Sr)(Al,Ta)O_3$; NAO - $NdAlO_3$. Values accompanied by asterisks (*) have been extracted from figures and should thus be considered approximate. These data are plotted in Fig. 3a. The *c*/*a* is calculated using $\frac{c}{a} = c/\sqrt{ab}$ when both *a* and *b* lattice parameters are available.

| **Authors** | **Ref.** | **Substrate** | **Film Type** | **Thickness (nm)** | ***a* (Å)** | ***b* (Å)** | ***c* (Å)** | **β (°)** | ***c/a*** | **Symmetry (method)** | **Tilting pattern** |
|---|---|---|---|---|---|---|---|---|---|---|---|
| Ederer & Spaldin | 25 | Calcs | - | - | 3.665 | 3.665 | 4.655 | 90 | 1.270 | *P4mm* | $a^0a^0c^0$ |
| Ricinschi *et al.* | 29 | STO | Mixed | 300 | 3.77 | | 4.65 | | | | |
| | | | | | 3.88 | | 4.07 | | | | |
| | | STO | Pure T | 300 | 3.72 | | 4.67 | | 1.255 | | |
| | | Calcs | - | - | 3.67 | | 4.64 | 90 | 1.264 | *P4mm* | $a^0a^0c^0$ |
| Ravindran *et al.* | 26 | Calcs | - | - | 3.7859 | | 4.8525 | 90 | 1.282 | *P4mm* | $a^0a^0c^0$ |
| Tutuncu *et al.* | 27 | Calcs LSDA | - | - | 3.70 | | 4.55 | 90 | 1.230 | *P4mm* | $a^0a^0c^0$ |
| | | Calcs GGA | - | - | 3.670 | | 4.639 | 90 | 1.264 | *P4mm* | $a^0a^0c^0$ |
| Bea *et al.* | 30 | LAO | Pure T | 7 | 3.79 | | 4.65(1) | | 1.232 | | |
| | | LAO | Pure T | 56 | 3.79 | | 4.65(1) | | 1.232 | $M_A$ (PFM) | |
| | | LAO | Pure T | 100 | 3.79 | | 4.65(1) | | 1.232 | $M_C$ ? (XRD) | |
| Ju *et al.* | 40 | Calcs | - | - | 3.689 | | 4.610 | 90 | 1.250 | *P4mm* | $a^0a^0c^0$ |
| Zeches *et al.* | 31 | LAO | Pure T | 17 | 3.83(8) | 3.77(2) | 4.62(7) | 88.67 | 1.216 | $M_C$ (XRD) | |
| | | LAO | Mixed | 53 | 3.84(4) | 3.75(3) | 4.64(4) | 88.51 | 1.223 | $M_C$ (XRD) | |
| | | LAO | Mixed | 89 | 3.66(7) | 3.58(6) | 4.69(1) | 88.63 | 1.296 | $M_C$ (XRD) | |
| | | LAO | Mixed | 120 | 3.66(5) | 3.59(1) | 4.70(1) | 88.79 | 1.297 | $M_C$ (XRD) | |
| | | YAO | Pure | ? | | | 4.667 | | | | |
| Hatt *et al.* | 41 | Calcs | - | - | | | | | | *Cc* | $a^-b^-c^0$ |
| Chen *et al.* | 42 | LSAO | Pure/Mixed | 20-210 | | | 4.651-4.674 | | | | |
| Mazumdar *et al.* | 43 | LAO | Pure T | 70 | 3.73 | | 4.65(2) | | 1.247 | $M_C$ (XRD) | |
| | | LAO | Pure T | 100 | 3.75 | | 4.65(2) | | 1.240 | | |
| Iliev *et al.* | 44 | LAO | Pure/Mixed | 70-200 | 3.79 | | 4.66 | | 1.230 | | |
| Chen *et al.* | 45 | LAO | Mixed | 70 | 3.818 | 3.740 | 4.662 | 88.12 | 1.234 | $M_C$ (XRD, PFM) | |
| | | | | | | | 3.99 | | | | |
| | (46) | LAO | Pure | | 3.811 | 3.734 | 4.670 | 88.2 | 1.238 | | |
| Chen *et al.* | 47 | LSAO | Mixed | 90 | 3.817 | 3.756 | 4.664 | 88.12 | 1.232 | $M_C$ (XRD, PFM) | |
| Dieguez *et al.* | 48 | Calcs | - | - | 3.75 | 3.75 | 4.745 | | 1.265 | *Pc* | |
| | | | | | 3.69 | 3.804 | 4.767 | | 1.272 | *Cm* | |
| | | | | | 3.758 | 3.758 | 4.726 | | 1.258 | $Pna2_1$ | |
| | | | | | 3.764 | 3.764 | 4.722 | | 1.255 | *Cc* | |
| | | | | | 3.707 | 3.707 | 4.763 | | 1.285 | *P4mm* | $a^0a^0c^0$ |
| He *et al.* | 49 | LAO | Mixed | ? | | | 4.64 | | | | |
| Liu *et al.* | 50 | STO | ? | 180 | 3.799 | | 4.673 | | 1.23 | | |
| | | | | | 3.82 | | 4.17 | | 1.092 | | |
| Nakamura *et al.* | 51 | LAO | Mixed/Pure T' (Co doped *x* = 0.075 to *x* = 0.3) | 260 | 3.800 | | 4.670-4.660 | | 1.229-1.226 | | |
| | | | | | 3.944-3.941 | | 3.944-3.942 | | | | |
| Christen *et al.* | 52 | LAO | Pure T' | 270 | 3.84(2) | 3.70(2) | 4.64(2) | 87.9(2) | 1.231 | $M_C$ (XRD) | |
| | | YAO | Mixed | 270 | 3.82(4) | 3.72(4) | 4.66(2) | 88.5(3) | 1.236 | $M_C$ (XRD) | |
| | | Calcs | - | - | a/b = 1 | | | 88.1 | | *Cc* ($M_A$) | $a^-b^-c^0$ |
| | | Calcs | - | - | a/b = 1 | | | 90 | | *Pm* ($M_C$) | $a^-b^0c^0$ |
| | | Calcs | - | - | a/b = 1.01 | | | 88.4 | | *P1* | $a^-b^-c^0$ |
| Bennett *et al.* | 53 | LAO | | | 3.735 | 3.735 | 4.651 | | 1.245 | | |
| | | LAO | Ba 10% | 90 | 3.781 | 3.781 | 4.627 | | | | |
| | | LSAO | Ba 10% | 90 | 3.823 | 3.823 | 4.532 | | | | |
| | | NGO | Ba 10% | 90 | 3.832 | 3.832 | 4.415 | | | | |
| | | | | | 3.832 | 3.832 | 4.224 | | | | |
| | | LSAT | Ba 10% | 90 | 3.851 | 3.851 | 4.380 | | | | |
| | | | | | 3.851 | 3.851 | 4.242 | | | | |
| Damodaran *et al.* | 54 | LAO | Pure T' | 28 | 3.74 | 3.82 | 4.649 | 88.1 | 1.23 | $M_C$ (XRD) | |
| | | | Mixed | 130 | 3.82 | 3.82 | 4.168 | | | | |
| Kreisel *et al.* | 55 | LAO | Mixed | 100 | 3.75(1) | 3.80(1) | 4.66(1) | 88.6 | 1.234 | | |
| | | | | | | | 4.19(2) | | | | |
| | | | | | | | 3.97(1) | | | | |
| Siemons *et al.* | 56 | LAO | Mixed 25 °C | 300 | 3.842 | 3.770 | 4.67(2) | 88.1(3) | 1.227 | | |
| | | | | | | | 4.423 | | | | |
| Chen *et al.* | 57 | LAO | Pure T’ | 10 | | | 4.64 | | | | |
| | | LAO | Mixed | 80 | 3.811(1) | 3.734(2) | 4.670(2) | 88.23 | 1.239 | $M_C$ (XRD) | |
| | | | | | | | 3.97 | | | | |
| | | | Triclinic (S’) | | 3.911 | 3.822 | 4.178 | α = 89.47 | β = 89.91 | γ = 89.45 | |
| | | | Triclinic (T’) | | 3.816 | 3.720 | 4.682 | α = 88.49 | β = 89.78 | γ = 89.84 | |
| Qi *et al.* | 58 | LAO | Pure T’ | 10 | | | 4.64 | | | | |
| Liu *et al.* | 59 | Calcs | - | - | 3.64 | 3.64 | 4.82 | 90 | 1.324 | *P4mm* | $a^0a^0c^0$ |
| | | | | | 3.59 | 3.79 | 4.84 | 86.05 | 1.312 | *Pm* | |
| Chen *et al.* | 46 | LAO | Sm 10%; pure T’ | 32 | 3.782(4) | 3.764(4) | 4.653(5) | 89.4(1) | 1.233 | $M_C$ | |
| Infante *et al.* | 38 | LAO | Mixed | 70 | 3.79 | | 4.66 | | 1.230 | | |
| | | | | | 3.91 | | 4.10 | | | | |
| Ko *et al.* | 39 | LAO | Pure 30 °C | 30-40 | 3.82 | 3.74 | 4.64 | 88.01 | 1.227 | $M_C$ (XRD) | |
| | | | Pure 300 °C | 30-40 | 3.79 | 3.79 | 4.66 | 87.93 | 1.230 | $M_A$ (XRD) | |
| Liu *et al.* | 60 | STO | ‘Pure’ | 180 | 3.779 | | 4.677 | | 1.238 | | |
| Damodaran *et al.* | 61 | LAO | Pure | 30 | | | 4.628 | | | | |
| | | LAO | Mixed | 140 | | | 4.675 | | | | |
| | | | | | | | 4.383 | | | | |
| | | | | | | | 4.164 | | | | |
| | | LAO | Mixed | 250 | | | 4.677 | | | | |
| | | | | | | | 4.393 | | | | |
| | | | | | | | 4.156 | | | | |
| Lu *et al.* | 62 | LAO | Pure | 19 | 3.837 | 3.772 | 4.643 | 88.03 | 1.22 | | |
| | | LAO | | 38 | 3.837 | 3.757 | 4.651 | 88.04 | 1.225 | | |
| | | LAO | | 57 | 3.834 | 3.745 | 4.655 | 88.07 | 1.228 | | |
| | | LAO | | 86 | 3.821 | 3.743 | 4.667 | 88.16 | 1.234 | | |
| | | LAO | | 114 | 3.81 | 3.72 | 4.667 | 88.17 | 1.240 | | |
| Liu *et al.* | 63 | LAO | Pure/Mixed | 20-165 | | | 4.631-4.652* | | | | |
| | | | ? | ? | 3.81 | 3.76 | 4.64 | 88.5 | 1.226 | | |
| Woo *et al.* | 64 | LAO | Pure/Mixed | 18-87 | | | 4.637-4.657* | | | | |
| | | NAO | Pure | 14.4-90.2 | | | 4.706-4.699* | | | | |
| Liu *et al.* | 65 | YAO | Pure T’ | 18 | 3.79 | 3.75 | 4.63 | 88.83 | 1.228 | | |
| Haislmaier *et al.* | 66 | YAO | Pure T’ | 25 | 3.751 | 3.751 | 4.67 | | 1.255 | | |
| Kim *et al.* | 67 | LAO | La-doped 5% | | 3.80 | 3.76 | 4.66 | 88.3 | 1.233 | | |
| Zhao *et al.* | 68 | Sapphire | Pure T | 60 | 3.77 | 3.77 | 4.65 | | 1.233 | | |
| Dixit *et al.* | 69 | Calcs R’ | - | - | 7.80 | 7.80 | 7.80 | | | | $a^-a^-a^-$ |
| | | Calcs S’ | | | 3.80 | 3.703 | 4.149 | | 1.092 | | $a^-b^-c^-$ |
| | | Calcs T’ | | | 3.75 | 3.671 | 4.596 | 88.1 | 1.226 | | $a^-b^-c^0$ |

**Table II**. Summary of the various phases discussed in this review.

| Phase | Lattice Parameters (Å) | | | *c*/*a* | Tilt of *c** axis (°) | Symmetry | Conditions for formation |
|---|---|---|---|---|---|---|---|
| | a | b | c | | | | |
| T’ | 3.71-3.84 | 3.69-3.78 | 4.62-4.71 | 1.2-1.3 | none | $M_C$ | Strain; $Bi_2O_3$ |
| R’ | ~3.96 | ~3.96 | 3.96-3.99 | ~1.01 | none | $M_A$ ? | Even thicker films |
| $S'_{tilt}$ | 3.82-3.92 | 3.82-3.92 | 4.16-4.19 | ~1.07 | 2.5-3 | $M_A$ ? | $T'_{tilt}$; thicker films |
| $T'_{tilt}$ | 3.71-3.84 | 3.69-3.78 | 4.62-4.71 | 1.2-1.3 | 1.5-2 | $M_C$ / $M_A$ ? | $S'_{tilt}$; thicker films |
| $S_2$’ | ? | ? | ? | ? | ? | ? | Interface between $S'_{tilt}$ and $T'_{tilt}$ |

## 2. Structure and symmetry

### 2.1 Structure of the various phases, with a focus on the intermediate phase

Three principal structural variants are relevant to our discussion of mixed-phase $BiFeO_3$ thin films. These structures are illustrated in Fig. 1, and their physical properties summarised in Table II (a more complete picture can be drawn from Table I). Following the notation introduced by Beekman *et al.*[70], we use **T'**, **R'**, and **S'$_{tilt}$** to refer to the metastable phase with large tetragonality, the bulk-derived rhombohedral-like phase, and an intermediate phase with its *c**-axis strongly tilted[d] from the substrate normal, respectively. A variant of the T' phase with its *c**-axis tilted from the substrate normal is denoted **T'$_{tilt}$**. In addition, we use **S$_2$'** to refer to a peculiar phase that appears in x-ray diffraction (XRD) measurements as a diffuse peak or 'tail' near the T' phase peak. Remarkably, even after many years, this diffuse peak – which appears in almost all published XRD data – has received no attention. Here we consider its origin, and discuss whether it plays a role in the fascinating physical behaviour observed in mixed phase BFO thin films.

Figure 3 presents the out-of-plane lattice constants of BFO films as a function of in-plane lattice constant in the T', S'$_{tilt}$ and R' phases, collated from published data (see Table I for the T' and S'$_{tilt}$ phases, Refs. [16,45,53,71,72] for the R' phase), along with our experimental data. With this large dataset, a least squares linear fit to the R' like phase data yields a Poisson ratio of **0.59 ± 0.07**, a value fairly typical for ferroelectric oxides. The data in Fig. 3 also suggest that T' BFO is a mechanically 'hard' material, evidenced by the negligible change in *c* lattice constant for in-plane lattice constants ranging from 3.72 to 3.80 Å, and the corresponding large volume change over this range. In contrast, the R' phase volume remains almost constant over a strain range of ~3.5 % relative to the bulk structure. These observations are consistent with first-principles calculations of the elastic properties of the T' and R' phases[73] in which it was found that the T' phase is easier to compress in the out-of-plane direction and with a lower bulk modulus than the R' phase.

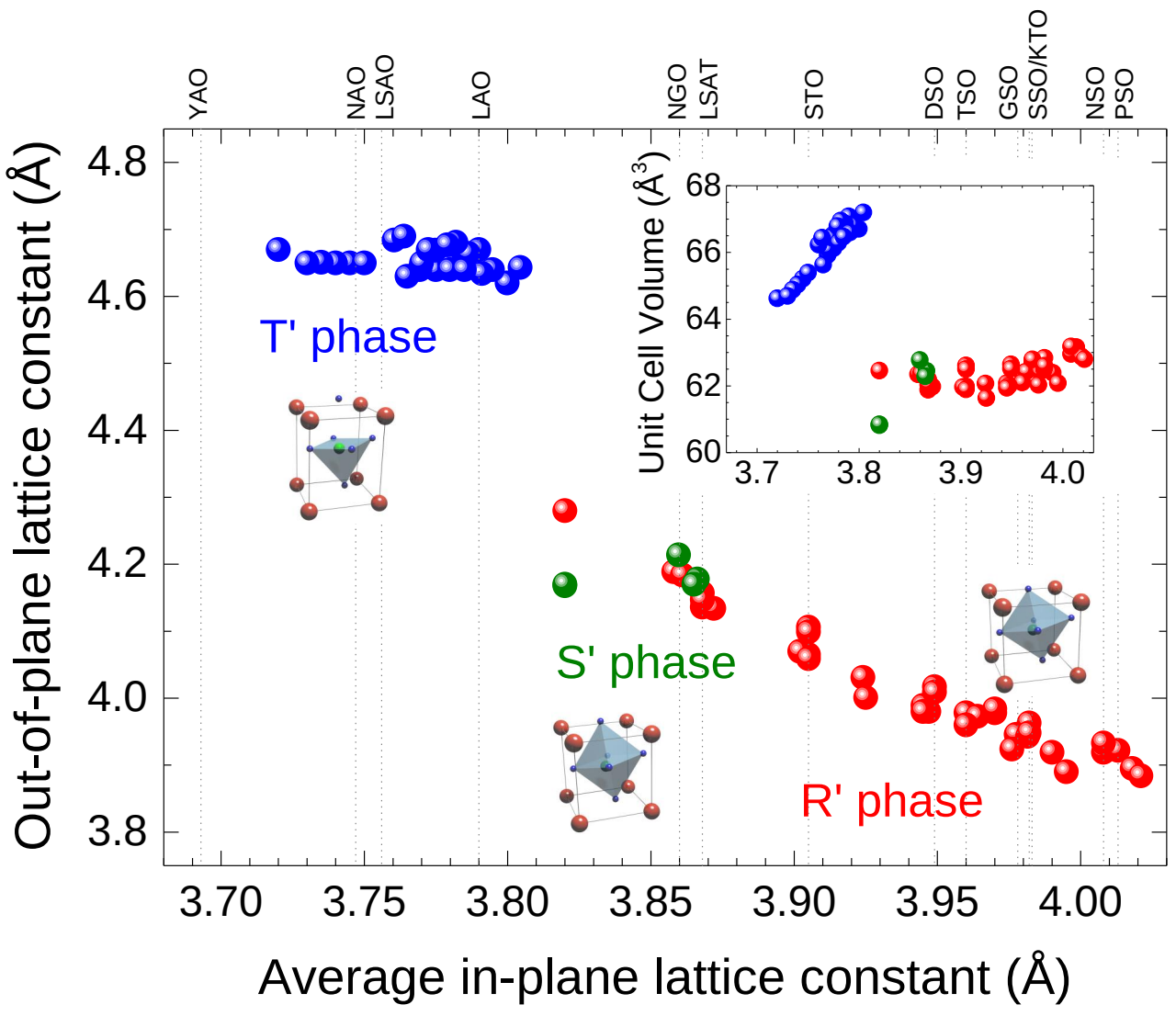

Figure 3. Lattice constants of R' and T' phase $BiFeO_3$ thin films, collated from published data (undoped BFO only). Inset: unit cell volume (assuming a simple tetragonal-like unit cell) as a function of in-plane lattice constant. The in-plane lattice constants of commonly-used single-crystal substrates are indicated in the main figure: YAO = $YaAlO_3$; NAO = $NdAlO_3$; LSAO = $LaSrAlO_4$;

---

[d] A note on nomenclature: to avoid confusion, in this manuscript we use the word 'tilt' to denote the *unit cell* tilt of the various phases, while the word 'rotation' is used to denote *octahedral* rotations/tilts.

NGO = $NdGaO_3$; DSO, TSO, GSO, SSO, NSO, PSO = $RScO_3$ where R = Dy, Tb, Gd, Sm, Nd, Pr respectively; KTO = $KTaO_3$. (Note that the plotted in-plane lattice constant is that measured for the BFO film, *not* the substrate lattice constant.)

The **T' phase** is a large axial ratio, super 'tetragonal' phase of BFO with $c/a$ ~ 1.23. Despite its common 'T-phase' description, its structure is generally *not* tetragonal but monoclinic, of either $M_A$ or $M_C$ symmetry (following standard notation[74]). This is the predominant phase in T' and mixed-phase films. The **R' phase** is an almost fully-relaxed rhombohedral-like phase (possibly of monoclinic symmetry, as in weakly-strained BFO films[75], but as yet not experimentally verified in mixed-phase films), which closely resembles the bulk form of BFO. The **S'$_{tilt}$ phase** is a strongly-compressively strained version of the R' phase, with its out-of-plane $c$*-axis tilted by up to ~3° to the film normal (see Fig. 4e). XRD reciprocal space maps (RSMs) around symmetric reflections such as 001 or 002 (Fig. 4b), along with atomic force microscopy (AFM) topography scans (Fig. 4c) serve to identify this tilted phase. The lattice constants of this phase (see Fig. 3) correspond to R' BFO under a compressive strain of around -2.6% (Refs. [16,17,45]). Finally, the **T'$_{tilt}$ phase** has almost identical lattice constants to the T' phase, but has its $c$*-axis tilted by around 1.5° to the film normal (Fig. 4e).

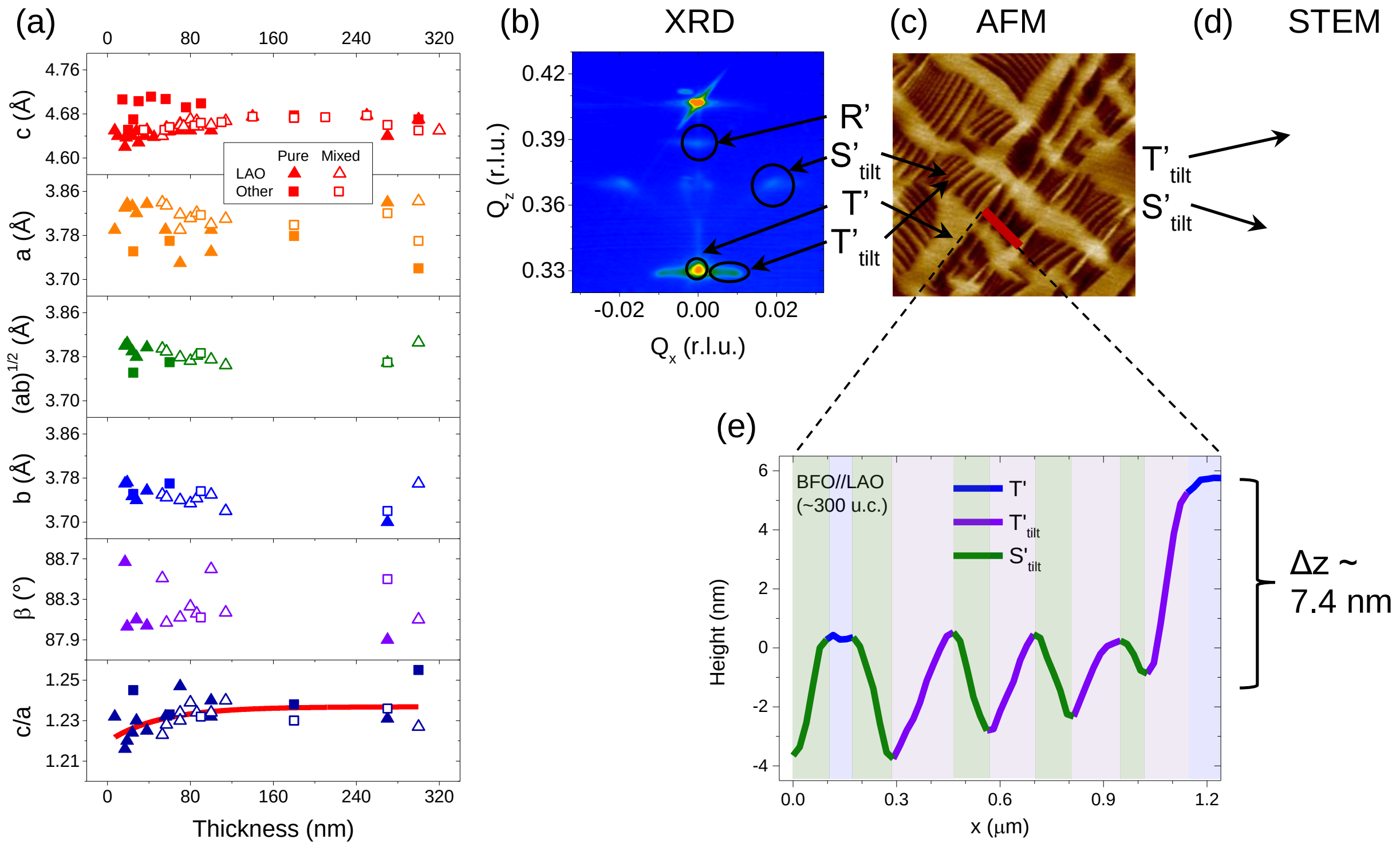

Figure 4. Structural properties of various $BiFeO_3$ phases formed in thin films (undoped BFO only). (a) Lattice parameters and monoclinic angle as a function of thickness (the raw data are presented in Table I). Filled and open symbols correspond to pure T' phase films and mixed phase films respectively. Triangles denote films grown on $LaAlO_3$ (LAO) substrates, while squares are for other substrates. (b-d) Experimental techniques for identifying the various phases: (b) XRD reciprocal space mapping around the 002 symmetric reflection; (c) Atomic force microscopy (AFM) topography (5 x 5 μm$^2$ scan); (d) Scanning transmission electron microscopy (STEM) (from Ref. [81]). (e) Line profile of the AFM scan indicated by the red line in (c), exemplifying the different tilting angles and height difference of the various phases. (d) reprinted from Ref. [81]. Copyright 2012 American Physical Society.

Published structural parameters of the T' phase in pure and mixed-phase films as a function of thickness are presented in Fig. 4a (the data for which can be found in Table I). These data can be loosely categorised into three regimes: for thin films (< 30-50 nm), the *c*/*a* is ~1.225; here the epitaxial misfit strain energy is sufficient to stabilise a pure T' phase, albeit 'strained' (probably tensile, relative to the 'stable' T' phase[41,76]). In the intermediate thickness regime (~60–200 nm), strain relaxation occurs and the metastable T' phase becomes less favourable than more stable bulk-derived polymorphs. The $S'_{tilt}$ and R' phases thus form, giving rise to mixed-phase regions (Figs. 4c,e). These distinctive 'striped' features comprise alternating $S'_{tilt}$ and $T'_{tilt}$ regions, and appear as depressions in the surface, as shown in Fig. 4e, since the average lattice constant of these mixed phases is lower. A quick calculation using the film thickness (~300 unit cells), out-of-plane lattice parameter of T' (4.66 Å), and average of the $T'_{tilt}$ and $S'_{tilt}$ lattice parameters (4.42 Å), yields a height difference of ~7.2 nm between mixed regions and pure T', as exemplified in Fig. 4e. The tilting observed in the $S'_{tilt}$ and $T'_{tilt}$ phases (Fig. 3e) is possibly due to shear strain induced during the relaxation process. For thicker films (> 200 nm), various groups have reported a pure T' phase, where substrate choice and possibly other growth conditions (e.g. bismuth excess, to which we return in Section 3) can favour the metastable T' phase, but thick films typically exhibit a mixture of the T', $S'_{tilt}$, and R' phases (in varying proportions, depending on substrate and growth conditions).

The out-of-plane lattice constant and *c*/*a* ratio of the T' phase, somewhat oddly, appear to *increase* with thickness, but *only* in the presence of the tilted mixed-phase regions. It may thus be possible that the presence of the $S'_{tilt}$ phase allows the T' and $T'_{tilt}$ phases to approach their 'natural' parameters, thereby decreasing (increasing) their in-plane (out-of-plane) constant. Indeed, when the thickness is increased further (above around 150 nm), the R' phase becomes more dominant and the T' phase then shrinks again in the out of plane direction (Fig. 4a, panel 1). It is interesting to note that there is one set of data that appears to go against this trend (the upper values of *c* in the thickness range of 10-90 nm in Fig 4a). These data correspond to films grown on $NdAlO_3$ (NAO) substrates[64], where curiously the relaxation mechanism is observed to be opposite to that which occurs for BFO films on LAO substrates. Moreover, the mixed-phase regions typically observed in thicker films are suppressed when using the NAO substrate, pointing to a unusual mechanism for stabilising the T' phase to larger thicknesses[64]. Returning now to the general case, further increases in thickness beyond ~200 nm typically then simply destroy the T' phase, resulting in the R' phase only, and finally, in very thick films (~500 nm) the bulk rhombohedral phase can be recovered[61]. Finally, we point out that Fig. 4a includes structural data of T' phases stabilised by mechanisms other than epitaxial strain, on non-lattice-matched substrates such as $SrTiO_3$ (Ref. [60]) and $NdScO_3$ (Sando, unpublished data), with comparable structural parameters, attesting to the true metastable nature of this phase.

The monoclinic angle β (Fig. 4a, panel 5) appears not to exhibit a clear trend with thickness, although a weak decrease in distortion of the unit cell (increase in β) with increasing thickness may be inferred. The angle β reported in the literature can be grouped into two different value ranges, the most prevalent being β ~ 88.1°, with a few reports finding β ~ 88.5°. The existence of these two types of distortion may be related to various phases that are close in energy[48] with distortions of different directions, types, and magnitudes.

We turn our attention now to the **$S_2$' phase**, evident from the diffuse peak above the main T' peak in the RSM of Fig 4b. A large number of papers present XRD data which evidence this peak (see for example Refs. [42,45,54–57,61,62,65,77–79]), but the discussion of its origin or characteristics is scarce. In the work of Siemons *et al.*[56,80], at temperatures above 175 °C this phase disappears, leading these authors to conclude it is a 'secondary room-temperature phase'. The existence of this phase could be related to the crystal symmetry: at room temperature the structure is $M_C$, while at temperatures above ~100 °C the symmetry of the T' phase is $M_A$. Temperature-dependent phase transitions are discussed in more detail in Section 4.

We conjecture that the origin of the $S_2$' phase peaks observed in XRD is indeed the strain gradient between the $T'_{tilt}$ and $S'_{tilt}$ phases. As the peak observed in XRD is typically very low in intensity and quite broad it cannot arise from a well-ordered crystalline bulk phase. Diffuse peaks such as these are often ascribed either to very local disordered phases, or to phases showing progressive changes in lattice parameter, leading to significant peak broadening. In mixed-phase BFO films, however, AFM topography images do not reveal the presence of secondary phase crystallites on the surface. In addition, the highest resolution aberration-corrected scanning-TEM (STEM) cross-section images (as illustrated in Fig. 4e, from Ref. [81]) show no evidence of secondary phases or short-range ordering. It is interesting to note that despite the significant difference in the *c*/*a* ratio (~13%) of the two phases, both are observed to coexist coherently at the interface[82] without the need to form structural defects[e]. Not only is this rather remarkable from an epitaxy/strain mismatch perspective, the change of in-plane (out-of-plane) lattice constant from ~3.75 Å to ~3.85 Å (4.65 Å to ~4.18 Å) over a few unit cells when traversing a T' → S' boundary gives rise to a large strain gradient (up to ~$10^7$ $m^{-1}$), forming an ideal platform for investigating flexoelectric effects[83,84]. The broad low-intensity XRD peak may in fact represent diffraction from the interfacial transitional regions wherein the large out-of-plane strain is gradually relaxed from the T' to S' phase. This is potentially one research problem to which state-of-the-art dark-field x-ray microscopy techniques[85] – which offer both high spatial and lattice resolution, *without* the need for destructive sample-preparation – could be very effectively applied.

### 2.2 Symmetry and oxygen octahedra rotation patterns

We now move to a discussion of the symmetry of the different phases. For monoclinic structures, four space groups are relevant to the T' phase of BFO: *Pm*, *Pc*, *Cm*, and *Cc* (Nos. 6-9 respectively). The first two correspond to the structure $M_C$, while the latter two can be described as $M_A$ (Ref. [74]). The difference in these structures lies in the direction of the shear distortion (and hence the polarization vector $\boldsymbol{P}$). For $M_C$, the unit cell is primitive (*i.e.* similar to the cubic perovskite cell) and $\boldsymbol{P}$ is tilted from the [001] direction toward the [100] direction (*i.e.*, $\boldsymbol{P}$ is constrained to the (010) plane). On the other hand, the $M_A$ phase has a base-centred unit cell which is doubled in volume and rotated by 45° relative to the primitive perovskite cell. The shear distortion is toward [110] and therefore $\boldsymbol{P}$ is constrained to lie in the (110) plane. It is important to note that the *Cm* and *Pm* space groups preclude oxygen octahedra rotations about more than one axis, whereas in *Pc* and *Cc* rotations are in general allowed.

Inspection of Table I indicates that most work finds that the symmetry of the T' phase is $M_C$. This information is typically extracted through inspection of the peak splitting in XRD reciprocal space maps[75] and/or domain wall orientation in PFM experiments[47]. Interestingly, in the seminal work of Béa *et al.*[30], while it was mentioned that PFM measurements indicated that the symmetry of their T' BFO was probably $M_A$, careful inspection of the peak shape in their RSM may in fact indicate that the structure was $M_C$.

[e] This observation may be related to the fact that the mixed-phase regions likely do not form *during* growth of the film, but rather *after* growth during the process of cooling to room temperature (Section 4.2).

The situation may actually be more involved: considering the mixed phase regions where the S'$_{tilt}$ and T'$_{tilt}$ regions are interspersed, Zhou *et al*. alluded to the T'$_{tilt}$ phase being closer to the $M_A$ structure through micro-XRD data[86]. This observation may be related to some requirement of the T'$_{tilt}$ phase to have similar symmetry to the S'$_{tilt}$ phase (which is most likely $M_A$ since it is a derivative of the rhombohedral bulk compound), in turn implying that the electric *E*-driven phase transition between these two variants may in fact be isosymmetric, as originally inferred[41]. Interestingly, at high temperatures (*i.e.* during film growth), the symmetry of the T' phase is $M_A$, implying that during cooling the polarization rotates, but perhaps not in the T'$_{tilt}$ regions.

An important issue in the mixed-phase BFO system relates to the octahedral rotation patterns – the *antiferrodistortive modes* – of the various phases. The antiferrodistortive modes in $ABO_3$ perovskites are extremely important, since these modes directly influence the structure[19], as well as the *B*-O-*B* bond angles and thus the superexchange and antisymmetric exchange interactions. In addition, since soft phonon modes, particularly near phase transitions, are associated with the rotation patterns of the octahedra, ascertaining the rotations could shed further light on temperature-, electric field-, and strain-driven phase transitions. Since an early report which observed the *absence* of octahedral rotations in the T' phase (*i.e.* a *Cm* structure)[87], no further x-ray characterisations have been reported. It is clear that the rotation patterns have been underexplored – as is immediately evident upon inspection of the rightmost column of Table I. Naturally, this deficiency can be attributed to the experimental challenges in characterising octahedral rotations, particularly when taking into account that these phases appear as nanoscale mixtures. That said, this issue should receive more focus, because if we can gain a strong understanding of the rotation patterns for each phase, this will allow us to determine the manner in which they are modified at the boundaries between the phases,[88] and in turn uncover, for instance, why the magnetic properties are so strongly modified in the strongly-compressed S' phase[49]. Moreover, octahedral rotations, manifest through rotorestrictive coupling terms[89], can have a very strong influence on, for instance, domain wall chirality, in turn affecting ferroic properties.

First-principles calculations regarding the T' phase of BFO generally find that octahedral rotations about the *c* axis are suppressed (in Glazer notation $a^-b^-c^0$) (Refs. [41,69]). The magnitude of the rotations about the in-plane directions is also expected to reduce considerably relative to the R' phase, as indicated in Fig. 5a. These observations can be understood by the two possible types of coupling between the polarization and oxygen octahedra rotations. To describe this in more detail, we introduce the $\omega_R$ vector which characterizes antiphase rotations, whose direction corresponds to the axis about which the oxygen octahedra rotate in an antiphase fashion, while its magnitude denotes the angle of these rotations.[90] The magnitude of the various rotations as a function of strain are plotted in Fig 5a, while the components of polarization over the same strain range are shown in Fig. 5b (these data are discussed in detail in Ref. [91]). As explored in Ref. [92], there is a strong *repulsive* bi-quadratic coupling between the *z*-component of the polarization and the *z*-component of $\omega_{R,}$, which explains why rotations about the *c* axis are suppressed when the T' phase (for which the polarization, *P*, has a large *z*-component) forms. However, there is also a second – less well known – coupling which involves the *x*- and *y*-components of the polarization as well as the *x*- and *y*-components of $\omega_{R,}$. This coupling is *collaborative* rather than repulsive, providing a successful explanation as to why the T' phase of BFO has in-plane components of polarization[93] *as well as* in-plane rotations. We point out also that these components cannot be too large, since (*i*) compressive strain, as a general rule, disfavours in-plane components of polarization and rotations (see Fig. 5b); and (*ii*) there are also repulsive bi-quadratic couplings between $P_x$ and $\omega_{R,x}$, as well as between $P_y$ and $\omega_{R,y}$.

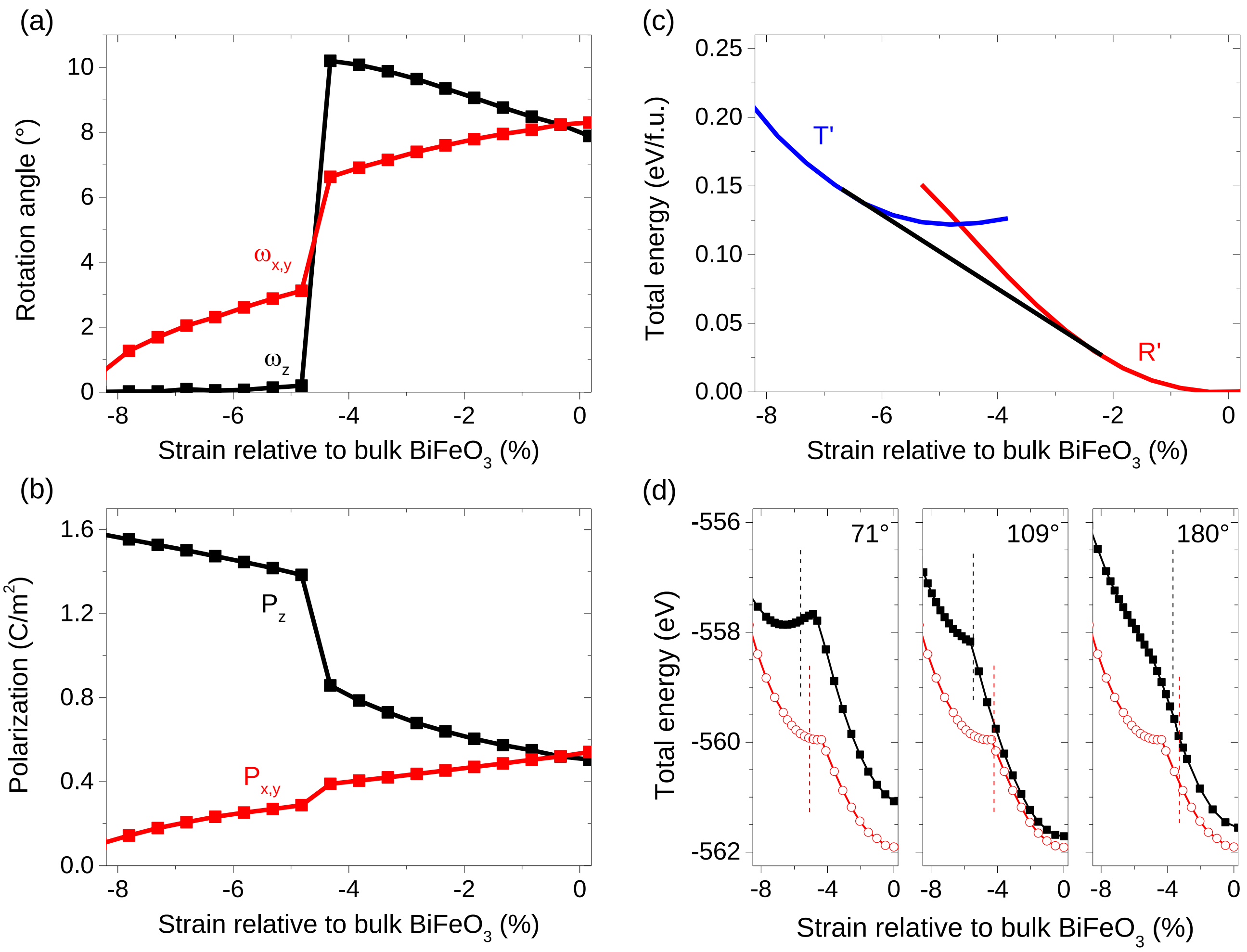

Figure 5. Calculated properties of R' and T' phases of BFO as a function of compressive strain (for simplicity the symmetry for both phases is *Cc*). (a) Octahedral rotations about the three axes, showing that *z* rotations are suppressed in the out-of-plane direction in the T' phase (from Ref. [91]). (b) Polarization in the in-plane and out-of-plane directions (from Ref. [91]). (c) Free energy of the two phases, showing why phase mixtures are observed for strains of -4 to -5%. (d) Energy of various types of domain walls in $BiFeO_3$ as a function of strain (from Ref. [122]).

### 3. The role of chemistry: Growth of T' phase and mixed phase BFO films

Stoichiometry appears to play a rather important role for the T' phase of BFO. Since this phase is metastable with its ground energy only slightly higher than that of the R' phase[25,26], the deviation of growth parameters from those optimal for R' phase stabilisation can be used to favour the formation of T' BFO. While this phase is typically formed by growing a thin film on a substrate with a lattice constant ~3.65-3.75 Å (strains of -7% to -4.5%), experiments suggest that *without* the use of strong compressive strain, a bismuth excess can stabilise the T' phase. For example, Ricinschi *et al.*[29], Béa[94], and Liu *et al.*[60] grew the T' phase on $SrTiO_3$, a substrate typically not expected to form the T' phase due to insufficient in-plane strain of only -1.6%. Even more remarkably, one of the present authors was able to stabilise the T' phase on $NdScO_3$, a substrate which would normally impart a *tensile* strain on the film (Sando, unpublished data). Interestingly, structural data (Fig. 6a-f) suggest that this polytype may in fact be a true tetragonal phase (due to the absence of peak splitting in the φ-scan and RSMs). This case of T' BFO on such a substrate was probably possible due to the formation of an interfacial defect layer (e.g. $Bi_2O_3$, Ref. [60]), allowing epitaxial matching between substrate and film. There have been other reports of apparently true tetragonal phases of BFO on various substrates such as sapphire[68] and $SrTiO_3$ (Ref. [60]).

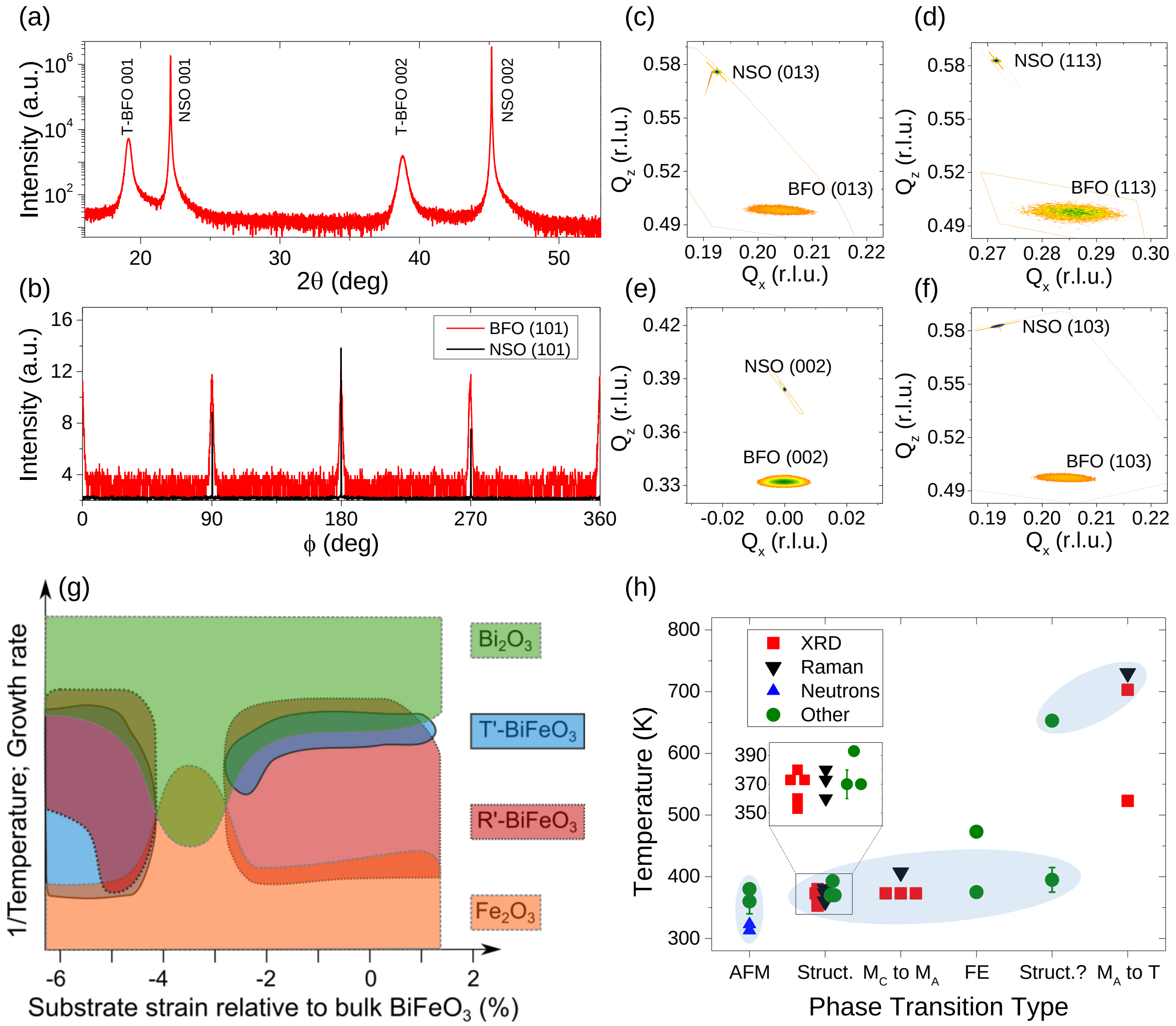

Figure 6. (a-f) X-ray diffraction data for a T' $BiFeO_3$ film on a $NdScO_3$ (NSO) substrate. The film appears to be phase pure (a), shows cube-on-cube epitaxy (b), shows no tilted phases (c), and is *not* strained to the substrate (d,e,f). Note also in (d,e,f) that the peak splitting (indicative of monoclinic structure) normally observed in T' BFO is *not* present. (g) The influence of bismuth stoichiometry: conceptual diagram highlighting phase stability regions for $BiFeO_3$ films under epitaxial strain. Note that BFO films generally do not form with a structure corresponding to ~-3.5% strain. T' BFO under low strain requires the presence of $Bi_2O_3$, while pure T' BFO can be obtained under strong compressive strain. (h) Reported phase transition temperatures from the literature. AFM = antiferromagnetic; Struct. = Structural; FE = ferroelectric; Struct? = possible structural. 'Other' denotes measurements by x-ray magnetic circular dichroism, transmission electron microscopy, second harmonic generation, piezoresponse force microscopy, or spectroscopic ellipsometry.

---

These observations highlight an important idea regarding the mindset of thin film growers. There has traditionally been somewhat an obsession with fabricating perfect (or ideal) epitaxial defect-free thin films, an approach that can overshadow subtle hints pointing to the potentially constructive role of non-stoichiometry and local chemical disorder. It may in fact be beneficial to upturn this mentality and rationalize how the influence of local defects can be harnessed for beneficial effects.

In these cases the T' phase formed upon either increasing the growth rate[60], or decreasing the substrate temperature[94,95], hinting at the presence of a bismuth excess, since these adjustments are expected to increase the bismuth content in the films. In some cases, a bismuth oxide phase is manifest in XRD data[29,60] or through transmission electron microscopy (TEM) (Ref. [96]). With this in mind, one may wonder what is the exact role of bismuth in forming the T' phase of BFO, *even* on appropriately lattice-matched substrates? For example, to stabilize thicker pure T' films on LAO (such as in Ref. [56]) the presence of $Bi_2O_3$ may be necessary. Further experiments, particularly TEM, should be performed on such samples in order to locate the bismuth rich phase, and thus elucidate its possible role in the stabilization of the T' phase to such large thicknesses.

An alternative mechanism responsible for the T' phase on non-lattice matched substrates was proposed by Ren *et al*.[95]. In this work, the authors grew pure and mixed phase T' films on STO by magnetron sputtering and suggested that the island growth mode, arising from the reduced adatom mobility at lower substrate temperatures, may have been able to induce sufficient strain to stabilise the T' phase.

Making use of these observations, we present a schematic phase diagram for the stabilisation of T' and/or R' BFO (Fig. 6g) as a function of strain and temperature/growth rate. This figure emphasises the role of temperature and/or growth rate along with the role of strain in stabilising the T' phase of BFO. For low strain, the ground state is R' BFO, but increasing growth rate or decreasing temperature allows to use bismuth oxides to stabilise the T' phase. For stronger compressive strain, the 'ground' state is T' BFO, but in thicker films a mixture of T' and R' phases can be formed. This is likely due to thermodynamic arguments: looking at Fig. 5c we see that the tangent line to the total energy curves explains why mixed T' and S' phases appear for average in-plane compressive strains of around 4-5 % (see Ref. [57] for more details). Since almost all experiments show that the BFO phase does not form with lattice constants in this range, we propose in our phase diagram that the BFO phase cannot exist here (note that the boundaries do not perfectly coincide between experiment and theory; due to for example the differences in computed lattice constants versus experimental values).

## 4. Phase transitions

One of the most exciting aspects of the mixed BFO ensemble is its extraordinary piezoelectric response[97]. This MPB-like behaviour in a lead-free compound has stimulated a mass of research, in XRD (structure), PFM (for domain-strain correlations) and AFM studies (to clarify the role of the various phases in the phase transitions). This remarkable response is the direct result of an electric field-induced phase transition, but the highly-strained BFO system also exhibits interesting temperature-driven ferroelectric and magnetic phase transitions not far above room temperature, bringing possibilities for enhanced magnetoelectric effects.

### 4.1 Electric field-induced phase transition

The electric field-induced T' → $S'_{tilt}$ phase transition and associated large structure change is the origin of the giant piezoelectric response in this material. A large number of groups have addressed this transition (see for instance Refs. [31,43,54,62,98,99]), predominantly using piezoresponse force microscopy (PFM) measurements where the ferroelectric domain structure and surface morphology before and after the transition can be probed. The general consensus is that upon the application of an electric field, the boundaries between the $T'_{tilt}$ and $S'_{tilt}$ phases are shifted, *i.e.* one phase is converted to the other. Since the two phases have strongly distinct structures, these transformations give rise to large strains. The coexistence of these phases is required to generate large piezoresponses, since the R' and T' phases separately have rather modest piezoelectric coefficients[100]. We point out here also that the concept of the coexistence of two rather different phases inside the same multi-domain structure resulting in remarkable enhancement of properties is rather general, since it has also been predicted to occur in domains comprising alternating *R*3*c* and *Pnma* phases in BFO (Ref. [101]).

The origin of this phase transition is likely a polarization rotation mechanism[54]. The complete evolution comprises the conversions $S'_{tilt}$ ↔ $T'_{tilt}$ ↔ T', and does not necessarily involve ferroelectric switching[54,98]. To intuitively understand this multistep transition, we consider the polarization direction (derived from Fig. 5b), as well as the tilting of the unit cells relative to the film normal (~2° for $S'_{tilt}$, ~1° for $T'_{tilt}$), for each of the phases. In the R' phase of BFO under low strain, the polarization points toward the <111> direction. The $S'_{tilt}$ phase, being a derivative of the R' phase, has polarization with a predominant <111> character, but rotated toward [001] due to strain[102] and the tilting of the structure. In the T' phase, the large axial ratio gives rise to polarization tilted by ~14° from the [001] direction, while in $T'_{tilt}$ the polarization is slightly rotated further in plane (~15° from [001]). As the polarization is progressively more out-of-plane in the $T'_{tilt}$ and then T' phase, an electric field applied to the $S'_{tilt}$ phase (such as in the mixed-phase regions) in the [001] direction can be used to favour the T' phase. The reverse transition is possible due to the low energy barrier that separates the T' and $S'_{tilt}$ phases[43,62], meaning that the phase conversion can be induced for applied field values lower than the coercive field for ferroelectric switching.

Interestingly, the switching mechanism may be related to a *second order* phase transition recently explored through first principles calculations (the strain-induced transition, in contrast, is of *first* order). These calculations indicate that in *bulk* BFO, strong electric fields applied along [001] can be used to access a T' phase with large axial ratio[103]. This recent observation highlights once again that the story of BFO is not all wrapped up and that theory has a significant role to play in unravelling the full story.

### 4.2 Temperature-driven phase transitions

Bulk BFO exhibits a range of phase transitions with temperature or upon application of hydrostatic pressure. The T' phase of this material has an equally diverse array of transitions, where the structure, magnetic order, and ferroelectric properties all undergo transitions just above room temperature[37,38,55,56,65,68,70,76,80,104–107]. These transitions are summarised in Fig. 6h. Upon increase from room temperature, the film undergoes an antiferromagnetic transition at 325 K (measured by neutron diffraction[37]) or 380 K (measured by Mössbauer spectroscopy[38]). In the same temperature window, a structural transition occurs (350–400 K), which has been shown to be from the $M_C$ to $M_A$ symmetry[56,65,70]. It is still not clear as to whether the structural and magnetic transitions are coupled; they may simply be coincidentally close. Further temperature increase induces a structural transition from the $M_A$ phase to a tetragonal *P*4*mm* phase[65,70]. This transition is typically measured by x-ray diffraction, but the two reported transition temperatures vary significantly (Refs. [65,70]). It is likely that this is the true ferroelectric transition (although switching ceases to be possible at lower temperatures[70]), so this discrepancy in measured transition temperature warrants further investigation.

The discrepancies in the measured transition temperatures may arise from different phases in the various samples cause by subtle changes in growth conditions. First-principles calculations have shown that different phases, each with different polarization directions, octahedral rotation patterns and polarization magnitudes, are close in energy.[48,52] In fact, it is rather remarkable that using different exchange-correlation functionals in density-functional calculations can have a dramatic effect on energy differences between some phases and even on the stability of some states in BFO (Refs. [48,69]).

## 5. Experimental challenges

In this section, we describe some of the unavoidable experimental challenges in the study of nanoscale mixtures such as this mixed-phase BFO system. The T' polymorph of BFO typically shows a resistivity values that are much lower than its R' phase counterpart. As a consequence, the predicted 150 μC/cm$^2$ polarization has never been experimentally confirmed in a pure T' phase film. While there is strong evidence[81,108] that the polarization is giant, it has not been reported unequivocally through a conventional polarization-electric field (*P*-*E*) hysteresis loop. Theoretical calculations[29] suggest that the T' polymorph of BFO should have a *lower* band gap than the R' phase, even though the *optical* band gap is 300-400 meV *higher* than the R' phase[109]. As discussed in Ref. [36], while the optical band gap of T' BFO is experimentally larger, the *electronic* band gap is smaller in T' BFO than in its rhombohedral counterpart. This lower electronic band gap, along with the possible existence of defect states in the gap[110], could in part be responsible for the higher leakage observed in T' phase specimens.

Another possible leakage mechanism may be related to the presence of bismuth oxides. It is known that bismuth oxide is a good ionic conductor[111] and as highlighted in Section 3, these oxides may be necessary (even in minuscule amounts) to facilitate the stabilisation of the T' phase. Indeed, it has been shown that trace levels of iron oxide impurities *not* easily detectable in laboratory-based XRD measurements can drastically alter the magnetic properties of BFO films[112]. Here we suggest that similarly, trace amounts of $Bi_2O_3$ may be present in T' films, enhancing the conductivity and hindering electrical property measurement.

Magnetic characterisation of antiferromagnetic thin films is generally a challenging task. Neutron diffraction, the method of most flexibility and choice, usually requires large samples (either powder or single crystal). The mixed phase films of BFO pose even further difficulties as the samples are typically rather thin. More local probes such as x-ray photoemission electron microscopy (PEEM), using x-ray magnetic linear/circular dichroism (XMLD, XMCD) have proven to be more applicable[35,49] to this particular system. However, significant challenges still remain – the spatial resolution of these techniques is not yet sufficient to pin down the magnetic moment in the phase boundaries, and complementary information regarding octahedral rotations (nano-XRD or similar) would be highly beneficial.

## 6. Conclusions and perspectives

From a physicist's perspective, the mixed-phase of bismuth ferrite is a veritable gold mine. When taking into account the multitude of fascinating qualities that exist in the parent compound (ferroelectricity, magnetic order, band gap in the visible range (~2.7 eV), magnetoelectric (ME) coupling, interesting domain walls, useful optical properties, etc.) and then placing the material 'on the brink', one can gain access to a very wide range of intriguing and useful physical phenomena.

In this review, we have focused on a number of key issues. Using a thorough analysis of the literature, we described the rather unusual trends in the structure with thickness, and pointed out that the crystal symmetry – in particular the octahedral rotations – in mixed-phase BFO warrants further experimental attention. We explained that growth conditions and bismuth stoichiometry is an interesting handle for controlling the phase fractions of these films, and we suggested that these factors may influence the electrical properties. The various temperature-driven phase transitions were summarised, and we highlighted their importance in the context of the 'multiferroic phase transition' proposed by some groups. It is also important to note here that enhanced ME coupling has not been reported at temperatures approaching this transition.

A short note on domain walls (DWs). Since the discovery of domain wall conductivity in BFO (Ref. [113]), and established now more generally in other multiferroics[114–116], so-called *domain wall nanoelectronics* has become a research area as rich and diverse as the interplay between lattice and functional properties. The now widely-accepted view is that DWs can possess properties which are entirely different from the domains which they separate[117,118], offering the exciting prospect of a system with a large DW density in which the ferroic properties are governed by the domain walls, rather than the domains. Domain wall engineering has recently emerged as one of the most intensely researched areas in oxide ferroelectrics[119–121]. In this context, it would be rather interesting to study properties of 180°, 71° and 109° DWs in BFO as a function of strain since the structure of the wall is strongly dependent on the strain magnitude. For small applied strains these domains consist of R' phases that have different <111> directions of polarization, while under large compressive strains the domains are made of different T' phases. Although theoretical predictions suggest the existence of a strain-driven change of hierarchy between domain wall energies (Fig. 5d) and unusual atomic arrangements at the domain walls[122], experimental investigations into this phenomenon are as yet lacking. This is an important issue, as strain tuning of domain walls could offer a highly systematic and controlled means for investigating the role of factors such as domain wall energies, magnitude of oxygen octahedra rotations, bonding symmetry, etc. on the unique electronic properties driven by domain walls.

Returning briefly now to the literature analysis of Fig. 2, it is interesting to note that in recent years the number of publications related to devices based on, or making use of, the tetragonal phase of BFO is increasing. Indeed, the large polarization of this phase is attractive for applications in ferroelectric tunnel junctions[123], and it can be used to induce interesting field effects in Mott insulators[124]. With ever-increasing interest in multiferroic and/or multifunctional devices, one can expect that in the future T' BFO will solidify its role in functional oxide thin film devices. We particularly anticipate continued evolution toward nano- or micro-devices, including in the realm of optics and electro-optics. The intriguing optical functionalities of bulk and R' BFO (such as photostriction[125], photovoltaic effects[126], and electro-optic effects[127]), in addition to probably further more exciting phenomena, are likely transferable to the T' phase system, and the specificities of the boundaries between the various phases, as well as the electric field-induce phase transitions, are likely to offer new avenues for multifunctional devices.

There are still a number of important issues and pending questions, a few of which we highlight here. An essential step for the progress in mixed-phase BFO study is the adaptation of nano-probes (XRD, XPS) to determine, on a local scale, the structural, electronic and magnetic properties of the various phases *in situ* (*i.e.* in the thin film geometry, not *after* TEM sample preparation which likely changes the strain state of the specimen[58]). Other intriguing possibilities could arise through the use of doping[128] and/or strain engineering[129] techniques to induce novel two-dimensional spin orders as proposed[130] by theorists. Even further, is it possible to use strain engineering to fabricate *pure* S' phase films (*i.e.* not in the intricate phase mixtures)? Is there a difference between the $S'_{tilt}$ phase and R' BFO grown on, e.g. LSAT substrates[16]? An important question to ask here is whether the remarkable properties exhibited by the $S'_{tilt}$ phase persist in pure form or if they are by their very nature a result of the mixed-phases around them. Theoretical approaches for the computed energy of the T' and R' phases as a function of strain are required, so as to provide guidance to experimentalists. The theoretical curves can then be used to simply explain why there is a mixing between T' and R' phases (as a result of thermodynamics, as suggested in Ref. [57]). Such calculations can also be used to pinpoint key findings; *e.g.* (i) why low-symmetry (*i.e.* triclinic) phases were predicted to exist in Ref. [57] (due to elastic mismatch between the phases), and (ii) consequences for magnetoelectricity when the R' phase becomes metastable[15,131].

More ambitiously, doping strategies *i.e.* chemical strain, could be utilised to create a *bulk* T' phase of BFO. Once again, first principles calculations have an important role to play in guiding further experimental efforts, so as to identify possible candidate dopants and the anticipated induced structural changes. Finally, while the influence of strain on polarization is well established[25,71,102], it would be certainly valuable to confirm experimentally the theoretical predictions of how the AFD vary with strain in the R' and T' phases, particularly with regard to how the $x$ and $y$ components of AFD change when increasing the magnitude of the compressive strains. This would also drive new thin film growth experiments focused on fine tuning phase and domain structures of such phases.

After a decade of research, the myriad facets of this system are still being revealed. We eagerly await the next instalment of the intriguing and stimulating mixed-phase bismuth ferrite story.

**Acknowledgements**

We wish to thank Stéphane Fusil, Yurong Yang, Brahim Dkhil, Manuel Bibes, and Claudio Cazorla for valuable

discussions. NV and DS acknowledge support from an ARC Discovery Grant. B.X. and L.B. thank the Department of Energy, Office of Basic Energy Sciences, under contract ER-46612.

## 7. References

[1] Z. Kutnjak, J. Petzelt, and R. Blinc, Nature **441**, 956 (2006).

[2] H. Fu and R.E. Cohen, Nature **403**, 281 (2000).

[3] L. Bellaiche, A. Garcia, and D. Vanderbilt, Phys. Rev. Lett. **84**, 5427 (2000).

[4] D. Dramjanovic, IEEE Trans. Ultrason. Ferroelectr. Freq. Control **56**, 1574 (2009).

[5] M. Ahart, M. Somayazulu, R.E. Cohen, P. Ganesh, P. Dera, H. Mao, R.J. Hemley, Y. Ren, P. Liermann, and Z. Wu, Nature **451**, 1 (2008).

[6] P. Mandal, A. Manjón-Sanz, A.J. Corkett, T.P. Comyn, K. Dawson, T. Stevenson, J. Bennett, L.F. Henrichs, A.J. Bell, E. Nishibori, M. Takata, M. Zanella, M.R. Dolgos, U. Adem, X. Wan, M.J. Pitcher, S. Romani, T.T. Tran, P.S. Halasyamani, J.B. Claridge, and M.J. Rosseinsky, Adv. Mater. **27**, (2015).

[7] W.J. Merz, Phys. Rev. B **78**, 52 (1950).

[8] K.J. Choi, M. Biegalski, Y. Li, A. Sharan, J. Schubert, R. Uecker, P. Reiche, Y. Chen, X. Pan, V. Gopalan, L.-Q. Chen, D.G. Schlom, and C.-B. Eom, Science **306**, 1005 (2004).

[9] J.H. Haeni, P. Irvin, W. Chang, R. Uecker, P. Reiche, Y.L. Li, S. Choudhury, W. Tian, M.E. Hawley, B. Craigo, A.K. Tagantsev, X.Q. Pan, S.K. Streiffer, L.Q. Chen, S.W. Kirchoefer, J. Levy, and D.G. Schlom, Nature **430**, 758 (2004).

[10] G. Catalan, A. Janssens, G. Rispens, S. Csiszar, O. Seeck, G. Rijnders, D.H.A. Blank, and B. Noheda, Phys. Rev. Lett. **96**, 1 (2006).

[11] G. Catalan and J.F. Scott, Adv. Mater. **21**, 2463 (2009).

[12] D. Sando, A. Barthélémy, and M. Bibes, J. Phys. Condens. Matter **26**, 473201 (2014).

[13] A.L. Roytburd, S.P. Alpay, V. Nagarajan, C.S. Ganpule, S. Aggarwal, E.D. Williams, and R. Ramesh, Phys. Rev. Lett. **85**, 190 (2000).

[14] Y. Yang, I.C. Infante, B. Dkhil, and L. Bellaiche, Comptes Rendus Phys. **16**, 193 (2015).

[15] J.C. Wojdeł and J. Íñiguez, Phys. Rev. Lett. **105**, 037208 (2010).

[16] I.C. Infante, S. Lisenkov, B. Dupé, M. Bibes, S. Fusil, E. Jacquet, G. Geneste, S. Petit, A. Courtial, J. Juraszek, L. Bellaiche, A. Barthélémy, and B. Dkhil, Phys. Rev. Lett. **105**, 057601 (2010).

[17] D. Sando, A. Agbelele, D. Rahmedov, J. Liu, P. Rovillain, C. Toulouse, I.C. Infante,  a P. Pyatakov, S. Fusil, E. Jacquet, C. Carrétéro, C. Deranlot, S. Lisenkov, D. Wang, J.-M. Le Breton, M. Cazayous, A. Sacuto, J. Juraszek,  a K. Zvezdin, L. Bellaiche, B. Dkhil, A. Barthélémy, and M. Bibes, Nat. Mater. **12**, 641 (2013).

[18] J.R. Teague, R. Gerson, and W.J. James, Solid State Commun. **8**, 1073 (1970).

[19] A.M. Glazer, Acta Crystallogr. Sect. B **28**, 3384 (1972).

[20] S.V. Kiselev, R.P. Ozerov, and G.S. Zhdanov, Sov. Phys. Dokl. **7**, 742 (1963).

[21] I. Sosnowska, T. Peterlin-Neumaier, and E. Steichele, J. Phys. C Solid State Phys. **15**, 4835 (1982).

[22] J. Wang, J. Neaton, H. Zheng, V. Nagarajan, S. Ogale, B. Liu, D. Viehland, V. Vaithyanathan, D. Schlom, U. Waghmare, N. Spaldin, K. Rabe, M. Wuttig, and R. Ramesh, Science **299**, 1719 (2003).

[23] N. Spaldin and M. Fiebig, Science **309**, 391 (2005).

[24] M. Fiebig, J. Phys. D. Appl. Phys. **38**, R123 (2005).

[25] C. Ederer and N.A. Spaldin, Phys. Rev. Lett. **95**, 257601 (2005).

[26] P. Ravindran, R. Vidya, A. Kjekshus, H. Fjellvåg, and O. Eriksson, Phys. Rev. B **74**, 224412 (2006).

[27] H.M. Tütüncü and G.P. Srivastava, Phys. Rev. B **78**, 1 (2008).

[28] K.Y. Yun, D. Ricinschi, T. Kanashima, M. Noda, and M. Okuyama, Jpn. J. Appl. Phys. **43**, L647 (2004).

[29] D. Ricinschi, K.-Y. Yun, and M. Okuyama, J. Phys. Condens. Matter **18**, L97 (2006).

[30] H. Béa, B. Dupé, S. Fusil, R. Mattana, E. Jacquet, B. Warot-Fonrose, F. Wilhelm, A. Rogalev, S. Petit, V. Cros, A. Anane, F. Petroff, K. Bouzehouane, G. Geneste, B. Dkhil, S. Lisenkov, I. Ponomareva, L. Bellaiche, M. Bibes, and A. Barthélémy, Phys. Rev. Lett. **102**, 217603 (2009).

[31] R.J. Zeches, M.D. Rossell, J.X. Zhang, a J. Hatt, Q. He, C.-H. Yang, A. Kumar, C.H. Wang, A. Melville, C. Adamo, G. Sheng, Y.-H. Chu, J.F. Ihlefeld, R. Erni, C. Ederer, V. Gopalan, L.Q. Chen, D.G. Schlom, N.A. Spaldin, L.W. Martin, and R. Ramesh, Science **326**, 977 (2009).

[32] D. Albrecht, S. Lisenkov, W. Ren, D. Rahmedov, I. a. Kornev, and L. Bellaiche, Phys. Rev. B **81**, 140401 (2010).

[33] C. Ederer and N.A. Spaldin, Phys. Rev. B **71**, 060401(R) (2005).

[34] L. Bellaiche, Z. Gui, and I. a Kornev, J. Phys. Condens. Matter **24**, 312201 (2012).

[35] Y.-C. Chen, Q. He, F.-N. Chu, Y.-C. Huang, J.-W. Chen, W.-I. Liang, R.K. Vasudevan, V. Nagarajan, E. Arenholz, S. V Kalinin, and Y.-H. Chu, Adv. Mater. **24**, 3070 (2012).

[36] D. Sando, Y. Yang, V. Garcia, S. Fusil, E. Bousquet, C. Carrétéro, D. Dolfi, A. Barthélémy, P. Ghosez, L. Bellaiche, and M. Bibes, Submitted (n.d.).

[37] G.J. MacDougall, H.M. Christen, W. Siemons, M.D. Biegalski, J.L. Zarestky, S. Liang, E. Dagotto, and S.E. Nagler, Phys. Rev. B **85**, 100406(R) (2012).

[38] I.C. Infante, J. Juraszek, S. Fusil, B. Dupé, P. Gemeiner, O. Diéguez, F. Pailloux, S. Jouen, E. Jacquet, G. Geneste, J. Pacaud, J. Íñiguez, L. Bellaiche, A. Barthélémy, B. Dkhil, and M. Bibes, Phys. Rev. Lett. **107**, 237601 (2011).

[39] K.-T. Ko, M.H. Jung, Q. He, J.H. Lee, C.S. Woo, K. Chu, J. Seidel, B.-G. Jeon, Y.S. Oh, K.H. Kim, W.-I. Liang, H.-J. Chen, Y.-H. Chu, Y.H. Jeong, R. Ramesh, J.-H. Park, and C.-H. Yang, Nat. Commun. **2**, 567 (2011).

[40] S. Ju, T.-Y. Cai, and G.-Y. Guo, J. Chem. Phys. **130**, 214708 (2009).

[41] A.J. Hatt, N.A. Spaldin, and C. Ederer, Phys. Rev. B **81**, 054109 (2010).

[42] Z. Chen, L. You, C. Huang, Y. Qi, J. Wang, T. Sritharan, and L. Chen, Appl. Phys. Lett. **96**, 252903 (2010).

[43] D. Mazumdar, V. Shelke, M. Iliev, S. Jesse, A. Kumar, S. V Kalinin, A.P. Baddorf, and A. Gupta, Nano Lett. **10**, 2555 (2010).

[44] M.N. Iliev, M. V Abrashev, D. Mazumdar, V. Shelke, and A. Gupta, Phys. Rev. B **82**, 014107 (2010).

[45] Z. Chen, Z. Luo, C. Huang, Y. Qi, P. Yang, L. You, C. Hu, T. Wu, J. Wang, C. Gao, T. Sritharan, and L. Chen, Adv. Funct. Mater. **21**, 133 (2011).

[46] W. Chen, W. Ren, L. You, Y. Yang, Z. Chen, Y. Qi, X. Zou, J. Wang, T. Sritharan, P. Yang, L. Bellaiche, and L. Chen, Appl. Phys. Lett. **99**, 222904 (2011).

[47] Z. Chen, Z. Luo, Y. Qi, P. Yang, S. Wu, C. Huang, T. Wu, J. Wang, C. Gao, T. Sritharan, and L. Chen, Appl. Phys. Lett. **97**, 242903 (2010).

[48] O. Diéguez, O.E. González-Vázquez, J.C. Wojdeł, and J. Íñiguez, Phys. Rev. B **83**, 094105 (2011).

[49] Q. He, Y.-H. Chu, J.T. Heron, S.Y. Yang, W.I. Liang, C.Y. Kuo, H.J. Lin, P. Yu, C.W. Liang, R.J. Zeches, W.C. Kuo, J.Y. Juang, C.T. Chen, E. Arenholz, a Scholl, and R. Ramesh, Nat. Commun. **2**, 225 (2011).

[50] H. Liu, P. Yang, K. Yao, and J. Wang, Appl. Phys. Lett. **98**, 102902 (2011).

[51] Y. Nakamura, M. Kawai, M. Azuma, M. Kubota, M. Shimada, T. Aiba, and Y. Shimakawa, Jpn. J. Appl. Phys. **50**, 031505 (2011).

[52] H.M. Christen, J.H. Nam, H.S. Kim, A.J. Hatt, and N.A. Spaldin, Phys. Rev. B **83**, 144107 (2011).

[53] C.J.C. Bennett, H. Kim, M. Varela, M.D. Biegalski, D.H. Kim, D.P. Norton, H.M. Meyer, and H.M. Christen, **2**, 6 (2011).

[54] A.R. Damodaran, C. Liang, Q. He, C.-Y. Peng, L. Chang, Y.-H. Chu, and L.W. Martin, Adv. Mater. **23**, 3170 (2011).

[55] J. Kreisel, P. Jadhav, O. Chaix-Pluchery, M. Varela, N. Dix, F. Sánchez, and J. Fontcuberta, J. Phys. Condens. Matter **23**, 342202 (2011).

[56] W. Siemons, M.D. Biegalski, J.H. Nam, and H.M. Christen, Appl. Phys. Express **4**, 095801 (2011).

[57] Z. Chen, S. Prosandeev, Z.L. Luo, W. Ren, Y. Qi, C.W. Huang, L. You, C. Gao, I.A. Kornev, T. Wu, J. Wang, P. Yang, T. Sritharan, L. Bellaiche, and L. Chen, Phys. Rev. B **84**, 094116 (2011).

[58] Y. Qi, C. Huang, Z. Chen, Z. Luo, Y. Wang, J. Guo, T. White, J. Wang, C. Gao, T. Sritharan, and L. Chen, Appl. Phys. Lett. **99**, 132905 (2011).

[59] Y. Liu, L. Ni, Z. Ren, C. Song, and G. Han, Integr. Ferroelectr. **128**, 71 (2011).

[60] H. Liu, P. Yang, K. Yao, K.P. Ong, P. Wu, and J. Wang, Adv. Funct. Mater. **22**, 937 (2012).

[61] A.R. Damodaran, S. Lee, J. Karthik, S. MacLaren, and L.W. Martin, Phys. Rev. B **85**, 024113 (2012).

[62] C.-L. Lu, J.-M. Liu, and T. Wu, Front. Phys. **7**, 424 (2012).

[63] H.-J. Liu, C.-W. Liang, W.-I. Liang, H.-J. Chen, J.-C. Yang, C.-Y. Peng, G.-F. Wang, F.-N. Chu, Y.-C. Chen, H.-Y. Lee, L. Chang, S.-J. Lin, and Y.-H. Chu, Phys. Rev. B **85**, 014104 (2012).

[64] C.-S. Woo, J.H. Lee, K. Chu, B.-K. Jang, Y.-B. Kim, T.Y. Koo, P. Yang, Y. Qi, Z. Chen, L. Chen, H.C. Choi, J.H. Shim, and C.-H. Yang, Phys. Rev. B **86**, 054417 (2012).

[65] H.-J. Liu, H.-J. Chen, W.-I. Liang, C.-W. Liang, H.-Y. Lee, S.-J. Lin, and Y.-H. Chu, J. Appl. Phys. **112**, 052002 (2012).

[66] R.C. Haislmaier, N.J. Podraza, S. Denev, A. Melville, D.G. Schlom, and V. Gopalan, Appl. Phys. Lett. **103**, 031906 (2013).

[67] K.-E. Kim, B.-K. Jang, Y. Heo, J. Hong Lee, M. Jeong, J.Y. Lee, J. Seidel, and C.-H. Yang, NPG Asia Mater. **6**, e81 (2014).

[68] Y.J. Zhao, Z.G. Yin, X.W. Zhang, Z. Fu, B.J. Sun, J.X. Wang, and J.L. Wu, ACS Appl. Mater. Interfaces **6**, 2639 (2014).

[69] H. Dixit, C. Beekman, C.M. Schlepütz, W. Siemons, Y. Yang, N. Senabulya, R. Clarke, M. Chi, H.M. Christen, and V.R. Cooper, Adv. Sci. 1500041 (2015).

[70] C. Beekman, W. Siemons, T.Z. Ward, M. Chi, J. Howe, M.D. Biegalski, N. Balke, P. Maksymovych, A.K. Farrar, J.B. Romero, P. Gao, X.Q. Pan, D.A. Tenne, and H.M. Christen, Adv. Mater. **25**, 5561 (2013).

[71] H. Jang, S.H. Baek, D. Ortiz, C.M. Folkman, R.R. Das, Y.H. Chu, P. Shafer, J.X. Zhang, S. Choudhury, V. Vaithyanathan, Y.B. Chen, D.A. Felker, M.D. Biegalski, M.S. Rzchowski, X.Q. Pan, D.G. Schlom, L.Q. Chen, R. Ramesh, and C.B. Eom, Phys. Rev. Lett. **101**, 107602 (2008).

[72] D.H. Kim, H.N. Lee, M.D. Biegalski, and H.M. Christen, Appl. Phys. Lett. **92**, 012911 (2008).

[73] H. Dong, C. Chen, S. Wang, W. Duan, and J. Li, Appl. Phys. Lett. **102**, 182905 (2013).

[74] D. Vanderbilt and M.H. Cohen, Phys. Rev. B **63**, 094108 (2001).

[75] G. Xu, H. Hiraka, G. Shirane, J. Li, J. Wang, and D. Viehland, Appl. Phys. Lett. **86**, 182905 (2005).

[76] K.-Y. Choi, S.H. Do, P. Lemmens, D. Wulferding, C.S. Woo, J.H. Lee, K. Chu, and C.-H. Yang, Phys. Rev. B **84**, 132408 (2011).

[77] C.-J. Cheng, C. Lu, Z. Chen, L. You, L. Chen, J. Wang, and T. Wu, Appl. Phys. Lett. **98**, 242502 (2011).

[78] N. Dix, R. Muralidharan, M. Varela, J. Fontcuberta, and F. Sánchez, Appl. Phys. Lett. **100**, 122905 (2012).

[79] P.S.S.R. Krishnan, J. a. Aguiar, Q.M. Ramasse, D.M. Kepaptsoglou, W.-I. Liang, Y.-H. Chu, N.D. Browning, P. Munroe, and V. Nagarajan, J. Mater. Chem. C **3**, 1835 (2015).

[80] W. Siemons, G.J. MacDougall, a. a. Aczel, J.L. Zarestky, M.D. Biegalski, S. Liang, E. Dagotto, S.E. Nagler, and H.M. Christen, Appl. Phys. Lett. **101**, 212901 (2012).

[81] M.D. Rossell, R. Erni, M.P. Prange, J.-C. Idrobo, W. Luo, R.J. Zeches, S.T. Pantelides, and R. Ramesh, Phys. Rev. Lett. **108**, 047601 (2012).

[82] Z.L. Luo, H. Huang, H. Zhou, Z.H. Chen, Y. Yang, L. Wu, C. Zhu, H. Wang, M. Yang, S. Hu, H. Wen, X. Zhang, Z. Zhang, L. Chen, D.D. Fong, and C. Gao, Appl. Phys. Lett. **104**, 2 (2014).

[83] C.-E. Cheng, H.-J. Liu, F. Dinelli, Y.-C. Chen, C.-S. Chang, F.S.-S. Chien, and Y.-H. Chu, Sci. Rep. **5**, 8091 (2015).

[84] K. Chu, B. Jang, J.H. Sung, Y.A. Shin, E. Lee, K. Song, J.H. Lee, C. Woo, S.J. Kim, S. Choi, T.Y. Koo, Y. Kim, S. Oh, M. Jo, and C. Yang, Nat. Nanotechnol. (2015).

[85] H. Simons, A. King, W. Ludwig, C. Detlefs, W. Pantleon, S. Schmidt, I. Snigireva, A. Snigirev, and H.F. Poulsen, Nat. Commun. **6**, 6098 (2015).

[86] J. Zhou, M. Trassin, Q. He, N. Tamura, M. Kunz, C. Cheng, J. Zhang, W.-I. Liang, J. Seidel, C.-L. Hsin, and J. Wu, J. Appl. Phys. **112**, 064102 (2012).

[87] B. Dupé, I.C. Infante, G. Geneste, P.-E. Janolin, M. Bibes, A. Barthélémy, S. Lisenkov, L. Bellaiche, S. Ravy, and B. Dkhil, Phys. Rev. B **81**, 144128 (2010).

[88] A.Y. Borisevich, H.J. Chang, M. Huijben, M.P. Oxley, S. Okamoto, M.K. Niranjan, J.D. Burton, E.Y. Tsymbal, Y.H. Chu, P. Yu, R. Ramesh, S. V Kalinin, and S.J. Pennycook, Phys. Rev. Lett. **105**, 087204 (2010).

[89] A.N. Morozovska, E.A. Eliseev, M.D. Glinchuk, L.Q. Chen, and V. Gopalan, Phys. Rev. B **85**, 1 (2012).

[90] I. a. Kornev, L. Bellaiche, P.-E. Janolin, B. Dkhil, and E. Suard, Phys. Rev. Lett. **97**, 157601 (2006).

[91] Y. Yang, M. Stengel, W. Ren, X.H. Yan, and L. Bellaiche, Phys. Rev. B **86**, 144114 (2012).

[92] I.A. Kornev and L. Bellaiche, Phys. Rev. B **79**, 100105 (2009).

[93] Z. Chen, X. Zou, W. Ren, L. You, C. Huang, Y. Yang, P. Yang, J. Wang, T. Sritharan, L. Bellaiche, and L. Chen, Phys. Rev. B **86**, 235125 (2012).

[94] H. Béa, Croissance, Caracterisation and Integrations Dans Des Heterostructures de Films Minces Du Multiferroic BiFeO3, Universite Paris 6, n.d.

[95] P. Ren, S.K. Cho, P. Liu, L. You, X. Zou, B. Wang, J. Wang, and L. Wang, AIP Adv. **3**, 012110 (2013).

[96] S. Nakashima, H. Fujisawa, M. Kobune, M. Shimizu, and Y. Kotaka, **05**, 3 (n.d.).

[97] J. Zhang, X. Ke, G. Gou, J. Seidel, B. Xiang, P. Yu, W.-I. Liang, A.M. Minor, Y. Chu, G. Van Tendeloo, X. Ren, and R. Ramesh, Nat. Commun. **4**, 2 (2013).

[98] R. Vasudevan, Y. Liu, J. Li, and W. Liang, Nano Lett. **11**, 828 (2011).

[99] R.K. Vasudevan, M.B. Okatan, Y.Y. Liu, S. Jesse, J.-C. Yang, W.-I. Liang, Y.-H. Chu, J.Y. Li, S. V. Kalinin, and V. Nagarajan, Phys. Rev. B **88**, 020402(R) (2013).

[100] J.X. Zhang, B. Xiang, Q. He, J. Seidel, R.J. Zeches, P. Yu, S.Y. Yang, C.H. Wang, Y.-H. Chu, L.W. Martin, A.M. Minor, and R. Ramesh, Nat. Nanotechnol. **6**, 98 (2011).

[101] D. Wang, E.K.H. Salje, S.-B. Mi, C.-L. Jia, and L. Bellaiche, Phys. Rev. B **88**, 134107 (2013).

[102] D. Sando, A. Agbelele, C. Daumont, D. Rahmedov, W. Ren, I.C. Infante, S. Lisenkov, S. Prosandeev, S. Fusil, E. Jacquet, C. Carrétéro, S. Petit, M. Cazayous, J. Juraszek, J.-M. Le Breton, L. Bellaiche, B. Dkhil, A. Barthélémy, and M. Bibes, Trans. R. Soc. A **372**, 20120438 (2014).

[103] M. Stengel and J. Íñiguez, arXiv Prepr. 1510.02715 (2015).

[104] Y. Qi, Z. Chen, L. Wang, X. Han, J. Wang, T. Sritharan, and L. Chen, Appl. Phys. Lett. **100**, 022908 (2012).

[105] D.H. Kim and D. Lim, J. Korean Phys. Soc. **62**, 734 (2013).

[106] Z. Luo, Z. Chen, Y. Yang, H.-J. Liu, C. Huang, H. Huang, H. Wang, M.-M. Yang, C. Hu, G. Pan, W. Wen, X. Li, Q. He, T. Sritharan, Y.-H. Chu, L. Chen, and C. Gao, Phys. Rev. B **88**, 064103 (2013).

[107] A. Kumar, S. Denev, R.J. Zeches, E. Vlahos, N.J. Podraza, A. Melville, D.G. Schlom, R. Ramesh, and V. Gopalan, Appl. Phys. Lett. **97**, 112903 (2010).

[108] J.X. Zhang, Q. He, M. Trassin, W. Luo, D. Yi, M.D. Rossell, P. Yu, L. You, C.H. Wang, C.Y. Kuo, J.T. Heron, Z. Hu, R.J. Zeches, H.J. Lin, A. Tanaka, C.T. Chen, L.H. Tjeng, Y.-H. Chu, and R. Ramesh, Phys. Rev. Lett. **107**, 147602 (2011).

[109] P. Chen, N.J. Podraza, X.S. Xu, A. Melville, E. Vlahos, V. Gopalan, R. Ramesh, D.G. Schlom, and J.L. Musfeldt,

Appl. Phys. Lett. **96**, 131907 (2010).

[110] A.J. Hauser, J. Zhang, L. Mier, R.A. Ricciardo, P.M. Woodward, T.L. Gustafson, L.J. Brillson, and F.Y. Yang, Appl. Phys. Lett. **92**, 222901 (2008).

[111] T. Takahashi and H. Iwahara, Mater. Res. Bull. **13**, 1447 (1978).

[112] H. Béa, M. Bibes, S. Fusil, K. Bouzehouane, E. Jacquet, K. Rode, P. Bencok, and A. Barthélémy, Phys. Rev. B **74**, 020101 (2006).

[113] J. Seidel, L.W. Martin, Q. He, Q. Zhan, Y.-H. Chu, A. Rother, M.E. Hawkridge, P. Maksymovych, P. Yu, M. Gajek, N. Balke, S. V Kalinin, S. Gemming, F. Wang, G. Catalan, J.F. Scott, N.A. Spaldin, J. Orenstein, and R. Ramesh, Nat. Mater. **8**, 229 (2009).

[114] G. Catalan, J. Seidel, R. Ramesh, and J. Scott, Rev. Mod. Phys. **84**, 119 (2012).

[115] R.K. Vasudevan, W. Wu, J.R. Guest, A.P. Baddorf, A.N. Morozovska, E. a. Eliseev, N. Balke, V. Nagarajan, P. Maksymovych, and S. V. Kalinin, Adv. Funct. Mater. **23**, 2592 (2013).

[116] Y. Geng, H. Das, A.L. Wysocki, X. Wang, S.-W. Cheong, M. Mostovoy, C.J. Fennie, and W. Wu, Nat. Mater. **13**, 163 (2013).

[117] D. Meier, J. Seidel, A. Cano, K. Delaney, Y. Kumagai, M. Mostovoy, N.A. Spaldin, R. Ramesh, and M. Fiebig, Nat. Mater. **11**, 284 (2012).

[118] S. Farokhipoor, C. Magén, S. Venkatesan, J. Íñiguez, C.J.M. Daumont, D. Rubi, E. Snoeck, M. Mostovoy, C. de Graaf, A. Müller, M. Döblinger, C. Scheu, and B. Noheda, Nature **515**, 379 (2014).

[119] T. Sluka, A.K. Tagantsev, D. Damjanovic, M. Gureev, and N. Setter, Nat. Commun. **3**, 748 (2012).

[120] T. Sluka, A.K. Tagantsev, P. Bednyakov, and N. Setter, Nat. Commun. **4**, 1806 (2013).

[121] R.G.P. McQuaid, L.J. McGilly, P. Sharma, A. Gruverman, and J.M. Gregg, Nat. Commun. **2**, 404 (2011).

[122] W. Ren, Y. Yang, O. Diéguez, J. Íñiguez, N. Choudhury, and L. Bellaiche, Phys. Rev. Lett. **110**, 187601 (2013).

[123] S. Boyn, S. Girod, V. Garcia, S. Fusil, S. Xavier, C. Deranlot, H. Yamada, C. Carrétéro, E. Jacquet, M. Bibes, A. Barthélémy, and J. Grollier, Appl. Phys. Lett. **104**, 052909 (2014).

[124] H. Yamada, M. Marinova, P. Altuntas, A. Crassous, L. Bégon-Lours, S. Fusil, E. Jacquet, V. Garcia, K. Bouzehouane, A. Gloter, J.E. Villegas, A. Barthélémy, and M. Bibes, Sci. Rep. **3**, 2834 (2013).

[125] B. Kundys, M. Viret, D. Colson, and D.O. Kundys, Nat. Mater. **9**, 803 (2010).

[126] T. Choi, S. Lee, Y.J. Choi, V. Kiryukhin, and S.-W. Cheong, Science **324**, 63 (2009).

[127] D. Sando, P. Hermet, J. Allibe, J. Bourderionnet, S. Fusil, C. Carrétéro, E. Jacquet, J.-C. Mage, D. Dolfi, A. Barthélémy, P. Ghosez, and M. Bibes, Phys. Rev. B **89**, 195106 (2014).

[128] C.-H. Yang, D. Kan, I. Takeuchi, V. Nagarajan, and J. Seidel, Phys. Chem. Chem. Phys. **14**, 15953 (2012).

[129] D.G. Schlom, L.-Q. Chen, C.J. Fennie, V. Gopalan, D.A. Muller, X. Pan, R. Ramesh, and R. Uecker, MRS Bull. **39**, 118 (2014).

[130] C. Escorihuela-Sayalero, O. Diéguez, and J. Íñiguez, Phys. Rev. Lett. **109**, 247202 (2012).

[131] S. Prosandeev, I. a. Kornev, and L. Bellaiche, Phys. Rev. B **83**, 020102 (2011).

[132] Z. Fan, J. Xiao, H. Liu, P. Yang, Q. Ke, W. Ji, K. Yao, K.P. Ong, K. Zeng, and J. Wang, ACS Appl. Mater. Interfaces **7**, 2648 (2015).